\documentclass[12pt,a4paper]{article}

\usepackage{epsfig}
\usepackage{amsmath,amsfonts,amssymb}
\usepackage{cite}
\usepackage{colortbl}
\usepackage{pifont}

\newcommand{\gm}{\gamma^\mu}
\newcommand{\Wm}{W_{\mu}}
\newcommand{\Zm}{Z_{\mu}}

\providecommand{\openone}{\leavevmode\hbox{\small1\kern-3.8pt\normalsize1}}

\newcommand{\RE}{\operatorname{Re}}

\begin{document}

\begin{center}
\begin{Large}
{\bf Novel signatures for vector-like quarks}
\end{Large}

\vspace{0.5cm}
J.~A.~Aguilar--Saavedra$^a$, D.~E. L\'opez-Fogliani$^{b,c}$, C. Mu\~noz$^{d,e}$ \\[1mm]
\begin{small}
{$^a$ Departamento de F\'{\i}sica Te\'orica y del Cosmos, 
Universidad de Granada, \\ E-18071 Granada, Spain} \\ 
{$^b$ IFIBA, UBA \& CONICET, Departamento de F\'isica, FCEyN, Universidad de Buenos Aires, \\ 1428 Buenos Aires, Argentina} \\
{$^c$ Pontificia Universidad Cat\'olica Argentina, 1107 Buenos Aires, Argentina} \\
{$^d$ Departamento de F\'isica Te\'orica, Universidad Aut\'onoma de Madrid, Campus de Cantoblanco, E-28049 Madrid, Spain} \\
{$^e$ Instituto de F\'isica Te\'orica UAM-CSIC, Campus de Cantoblanco, E-28049 Madrid, Spain}
\end{small}
\end{center}

\begin{abstract}
We consider supersymmetric extensions of the standard model with a vector-like doublet $(T \, B)$ of quarks with charge $2/3$ and $-1/3$, respectively. Compared to non-supersymmetric models, there is a variety of new decay modes for the vector-like quarks, involving the extra scalars present in supersymmetry. The importance of these new modes, yielding multi-top, multi-bottom and also multi-Higgs signals, is highlighted by the analysis of several benchmark scenarios. We show how the triangles commonly used to represent the branching ratios of the `standard' decay modes of the vector-like quarks involving $W$, $Z$ or Higgs bosons can be generalised to include additional channels. We give an example by recasting the limits of a recent heavy quark search for this more general case.

\end{abstract}

\section{Introduction}

Vector-like quarks, whose left- and right-handed parts transform in the same representation of $\text{SU}(2)$, are usually considered in non-supersymmetric extensions of the standard model (SM) such as little Higgs~\cite{Perelstein:2003wd,Schmaltz:2005ky} and composite Higgs~\cite{Contino:2006qr,Contino:2006nn,Matsedonskyi:2012ym,Kaplan:1991dc} models. 
In supersymmetry, where a 
vector-like (Higgs doublet) representation already appears in the spectrum of the minimal supersymmetric standard model (MSSM)~\cite{Martin:1997ns} in order to cancel anomalies, vector-like quarks have been introduced mainly to raise the Higgs boson 
mass~\cite{Moroi:1991mg,Moroi:1992zk,Babu:2004xg,Babu:2008ge,Martin:2009bg,Martin:2010dc,Graham:2009gy,
Endo:2011mc,Martin:2012dg,Faroughy:2014oka,Ellis:2014dza,Lalak:2015xea,Nickel:2015dna,Collins:2015wua} and thus ameliorate the tension of the MSSM with the measured value $M_{h^0} = 125$ GeV.

Recently, two of the authors have suggested a reinterpretation of the Higgs doublet superfields, $\hat H_u$ and $\hat H_d$, as a fourth family of (vector-like) lepton superfields~\cite{Lopez-Fogliani:2017qzj} in the context of the
`$\mu$ from $\nu$' supersymmetric standard 
model ($\mu\nu$SSM)~\cite{LopezFogliani:2005yw}.
This seems to be more satisfactory from the theoretical viewpoint than the usual situation in supersymmetric models, where the Higgses are `disconnected' from the rest of the matter and do not have a three-fold replication.
In this framework, in analogy with the known first three families 
where for each lepton representation there is a quark counterpart, the possible existence of a vector-like quark doublet representation $(T \, B)$ was proposed in ref.~\cite{Lopez-Fogliani:2017qzj} as part of the fourth family.

The collider phenomenology of the production and decay of vector-like quarks in supersymmetric extensions of the SM can be quite different from the minimal vector-like extensions with a single Higgs doublet~\cite{AguilarSaavedra:2009es,Aguilar-Saavedra:2013qpa}. As already mentioned, supersymmetry  
requires the presence of two Higgs doublet superfields, 
with their scalar components $H_d$ and $H_u$ generating charged lepton and down-type quark masses, and up-type quark masses, respectively. Additional neutral singlet superfields can also exist, and their scalar components are in general mixed with the Higgses. For example, in the next-to-minimal supersymmetric standard model (NMSSM)~\cite{Ellwanger:2009dp} one extra singlet superfield $\hat N$ is included in order to solve the $\mu$ problem~\cite{Kim:1983dt}. 
In the $\mu\nu$SSM~\cite{LopezFogliani:2005yw,Munoz:2016vaa}, the $\mu$ problem is solved using three families of right-handed neutrino superfields $\hat \nu_L^c$, simultaneously reproducing at the tree level the correct neutrino physics~\cite{LopezFogliani:2005yw,Escudero:2008jg,Ghosh:2008yh,Bartl:2009an,Fidalgo:2009dm,Ghosh:2010zi,LopezFogliani:2010bf}. In this model, since $R$-parity is explicitly violated, all fields with the same quantum numbers mix together, and in particular the Higgses $H_u$ and $H_d$ turn out to be mixed with the right and left sneutrinos, $\widetilde \nu_R$ and $\widetilde \nu_L$, although the mixing with the left ones is very small and they are basically decoupled.

The additional scalars present in supersymmetry give new decay channels for the vector-like quarks, with distinctive signatures of multi-top, multi-bottom or even multi-Higgs signals. Current searches for heavy quarks $T$ and $B$ at the Large Hadron Collider (LHC) for heavy quark pair~\cite{Chatrchyan:2013uxa,Aad:2014efa,Khachatryan:2015axa,Aad:2015mba,Aad:2015gdg,Aad:2015kqa,
Khachatryan:2015gza,Khachatryan:2015oba} or single production~\cite{Khachatryan:2015mta,Aad:2015voa,Aad:2016qpo,Khachatryan:2016vph,Sirunyan:2017ezy,Sirunyan:2017tfc} focus on the standard decay modes,
\begin{align}
& T \to W^+ b \,,\quad T \to Z t \,, \quad T \to h^0 t \,, \notag \\
& B \to W^- t \,, \quad B \to Z b \,, \quad B \to h^0 b \,.
\label{ec:stddec}
\end{align}
The aim of this paper is to explore the additional signatures that can arise in models with non-minimal scalar sectors, 
using as benchmark the supersymmetric model with a vector-like quark doublet $(T \, B)$ proposed in ref.~\cite{Lopez-Fogliani:2017qzj}. Additional decay modes of vector-like quarks $T$ have been considered in composite Higgs~\cite{Serra:2015xfa}, little Higgs~\cite{Anandakrishnan:2015yfa} and two-Higgs doublet models~\cite{Arhrib:2016rlj}.

We begin by writing in section~\ref{sec:2} the interactions for a vector-like doublet extension when the scalar sector comprises two doublets $H_u$ and $H_d$, as in the case of the MSSM. This corresponds to a limit of negligible mixing of the neutral interaction eigenstates  $H_u^0$, $H_d^0$ with the additional scalars present in the benchmark model of ref.~\cite{Lopez-Fogliani:2017qzj}. In section~\ref{sec:3} we write the interactions in a more general scenario where $H_u^0$ and $H_d^0$ mix with a scalar singlet $\widetilde \nu_R$. We then study in section~\ref{sec:4} the decays of the heavy quarks $T$ and $B$  for the two scalar doublet model, and the model with two scalar doublets plus a singlet. There we analyse in particular the dependence on the model parameters of the $T$ and $B$ decay branching ratios for the standard and the new decay modes.

As we have remarked, current searches focus on the standard decay modes of the heavy quarks~(\ref{ec:stddec}). For those searches, we generalise in section~\ref{sec:5} the triangles that are commonly used by the ATLAS and CMS Collaborations, where they display the interpretation of their limits under the assumption that the branching ratios for the three modes~(\ref{ec:stddec}) add up to one, $\text{Br}(W) + \text{Br}(Z) + \text{Br}(h^0) = 1$. Relaxing such assumption, we will be able to plot these branching ratios within three-dimensional pyramids. Alternatively, a graphical representation by a set of equilateral triangles obtained by slicing the pyramids will be presented, and an example of how a standard search can be recast is given in section~\ref{sec:6}, where we show in a realistic case the resulting limits in this set of triangular slices. Finally, we discuss our results in section~\ref{sec:7}. Two appendices are devoted to collecting the partial widths for the different decay modes of the heavy quarks, and giving the relation between coordinates in the triangles and heavy quark decay branching ratios.


\section{Interactions for two scalar doublets}
\label{sec:2}

We consider a supersymmetric model with three SM quark generations $q_{Li} = (u^0_{Li} \; d^0_{Li})^T$, $u^0_{Ri}$, $d^0_{Ri}$, $i=1,2,3$, and an additional vector-like quark doublet $Q_{L,R} = (T^0 \; B^0)^T_{L,R}$. (We denote the weak eigenstates with zero superscripts.) The usual quark mass terms arise from Yukawa interactions with the scalar doublets $H_u = (H_u^+ \; H_u^0)^T$ and $H_d = (H_d^0 \; H_d^-)^T$. In the notation of four-component spinors, they are
\begin{equation}
\mathcal{L} = - y_{ij}^{u\,*} \, \bar q_{Li} u_{Rj} \epsilon H_u^* + y_{ij}^{d \, *} \, \bar q_{Li} d_{Rj} \epsilon H_d^* - y_{4j}^{u \, *} \, \bar Q_{L} u_{Rj} \epsilon H_u^* + y_{4j}^{d \, *} \, \bar Q_{L} d_{Rj} \epsilon H_d^* + \text{H.c.} \,,
\end{equation}
with $\epsilon = i \sigma^2$ the $2 \times 2$ anti-symmetric tensor. There is also a Yukawa interaction of the vector-like quark doublet with the scalar singlet $\widetilde \nu_R$~\cite{Lopez-Fogliani:2017qzj},
\begin{equation}
\mathcal{L} = - y_{44}^* \, \bar Q_L Q_R \widetilde \nu_R + \text{H.c.}\,, 
\label{ec:lY2}
\end{equation}
where we have applied a phase redefinition of the $B_R^0$
field to recover the conventions for the non-supersymmetric SM extensions with vector-like quarks~\cite{Aguilar-Saavedra:2013qpa}, which we use in the following.

After the neutral scalars acquire vacuum expectation values $\langle H_u^0 \rangle = v_u/\sqrt 2$, $\langle H_d^0 \rangle = v_d/\sqrt 2$, $\langle \widetilde \nu_R \rangle = v_R /\sqrt 2$,  the quark mass matrices are 
\begin{eqnarray}
\mathcal{L}_\text{mass} & = & - \left(\! \begin{array}{cc} \bar u_{Li}^0 & \bar T_L^0 \end{array} \!\right)
\left(\! \begin{array}{cc} y_{ij}^{u\,*} \frac{v_u}{\sqrt 2} & 0  \\ y_{4j}^{u\,*} \frac{v_u}{\sqrt 2} & y_{44}^* \frac{v_R}{\sqrt 2} \end{array} \!\right)
\left(\! \begin{array}{c} u^0_{Rj} \\ T^0_R \end{array}
\!\right) \notag \\
& & - \left(\! \begin{array}{cc} \bar d_{Li}^0 & \bar B_L^0 \end{array} \!\right)
\left(\! \begin{array}{cc} y_{ij}^{d\,*} \frac{v_d}{\sqrt 2} & 0 \\ y_{4j}^{d\,*} \frac{v_d}{\sqrt 2} &  y_{44}^* \frac{v_R}{\sqrt 2} \end{array} \!\right)
\left(\! \begin{array}{c} d^0_{Rj} \\ B^0_R \end{array}
\!\right) +\text{H.c.}
\label{ec:Lmass}
\end{eqnarray}

We assume that the new vector-like doublet eigenstates dominantly mix with the third generation, as it is expected from the mass hierarchy~\cite{Aguilar-Saavedra:2013wba}. (This assumption is also in agreement with stringent experimental constraints arising from flavour-changing processes at low energies~\cite{Barger:1995dd,Frampton:1999xi,Barenboim:2001fd,AguilarSaavedra:2002kr}.) Therefore, we can ignore the first two generations and write the relation between weak eigenstates and mass eigenstates as
\begin{align}
& \left(\!\! \begin{array}{c} t_{L,R} \\ T_{L,R} \end{array}  \!\!\right) =
U_{L,R}^u \left(\!\! \begin{array}{c} t^0_{L,R} \\ T^0_{L,R} \end{array} \!\!\right)
= \left(\! \begin{array}{cc} \cos \theta_{L,R}^u & -\sin \theta_{L,R}^u e^{i \phi_u} \\ \sin \theta_{L,R}^u e^{-i \phi_u} & \cos \theta_{L,R}^u \end{array}
\!\right)
\left(\!\! \begin{array}{c} t^0_{L,R} \\ T^0_{L,R} \end{array} \!\!\right) \,, \notag \\[2mm]
& \left(\!\! \begin{array}{c} b_{L,R} \\ B_{L,R} \end{array} \!\!\right)
= U_{L,R}^d \left(\!\! \begin{array}{c} b^0_{L,R} \\ B^0_{L,R} \end{array} \!\!\right)
= \left(\! \begin{array}{cc} \cos \theta_{L,R}^d & -\sin \theta_{L,R}^d e^{i \phi_d} \\ \sin \theta_{L,R}^d e^{-i \phi_d} & \cos \theta_{L,R}^d \end{array}
\!\right)
\left(\!\! \begin{array}{c} b^0_{L,R} \\ B^0_{L,R} \end{array} \!\!\right) \,,
\label{ec:mix}
\end{align}
with $t^0_{L,R} \equiv u^0_{L3,R3}$, $b^0_{L,R} \equiv d^0_{L3,R3}$. The mixing angles of left- and right-handed fields are not independent, but they satisfy~\cite{Dawson:2012di,Fajfer:2013wca,Atre:2011ae}
\begin{equation}
\tan \theta_L^u = \frac{m_t}{m_T} \tan \theta_R^u \,, \quad
\tan \theta_L^d = \frac{m_b}{m_B} \tan \theta_R^d \,.
\end{equation}
In the following we abbreviate $s_L^u = \sin \theta_L^u$, $c_L^u = \cos \theta_L^u$, etc. The agreement with the precisely measured $S$ and $T$ parameters and $R_b$, $R_c$, $A_\text{FB}^{b}$, $A_\text{FB}^{c}$ at LEP~\cite{ALEPH:2005ab} requires that these angles are small. (Note that for small mixing the $(4,4)$ entries in the mass matrices are approximately the heavy quark masses.) We write in this section the interactions in the mass basis in a MSSM-like case where the neutral interaction eigenstates $H_u^0$, $H_d^0$ have small mixing with the other scalars. In this case we have
\begin{align}
& H_u^0 = \frac{1}{\sqrt 2} \left( \cos \alpha \, h^0 + \sin \alpha \, H_1^0 + i \sin \beta \, G^0 + i \cos \beta \, P_1^0 \right) \,, \notag \\
& H_d^0 = \frac{1}{\sqrt 2} \left( -\sin\alpha \, h^0 + \cos \alpha \, H_1^0 - i \cos \beta \, G^0 + i \sin \beta \, P_1^0 \right) \,,
\end{align}
with $h^0$ being the SM-like Higgs boson, $H_1^0$ a scalar, $P_1^0$ a pseudo-scalar and $G^0$ a Goldstone boson. As usual, we define $\tan \beta = v_u/v_d$, $v = (v_u^2 + v_d^2)^{1/2} = 246$ GeV, and $\alpha$ is the mixing angle between the two neutral scalars $h^0$ and $H_1^0$. For the charged scalars we have
\begin{align}
& H_u^+ = \sin \beta \, G^+ + \cos \beta \, H^+ \,, \notag \\
& H_d^+ = -\cos \beta \, G^+ + \sin \beta \, H^+ \,, 
\end{align}
with $H^+$ the physical charged scalar and $G^+$ the Goldstone boson.
Furthermore, we take the alignment limit $\beta - \alpha = \pi/2$, since a small misalignment has no phenomenological consequences on the heavy quark decays. We remark that the interactions written below are the same for a non-supersymmetric type-II two-Higgs doublet model in which one doublet $H_u$ couples to charge $2/3$ quarks and the other doublet $H_d$ couples to charge $-1/3$ quarks.


\subsection{Light-heavy interactions}

These interactions determine the decay of the heavy quarks. The interactions with the $W$ and $Z$ bosons are the same as in the minimal model with one Higgs doublet,
\begin{eqnarray}
\mathcal{L}_W & = & -\frac{g}{\sqrt 2} \left[ \bar T \gm \left( V_{Tb}^L P_L + V_{Tb}^R P_R \right) b 
+ \bar t \gm \left( V_{tB}^L P_L + V_{tB}^R P_R \right) B \right] \Wm^+ +\text{H.c.} \,, \notag \\[1mm]
\mathcal{L}_Z & = & -\frac{g}{2 c_W} \left[ \bar t \gm \left( X_{tT}^L P_L + X_{tT}^R P_R \right) T
- \bar b \gm \left( X_{bB}^L P_L + X_{bB}^R P_R \right) B \right] \Zm +\text{H.c.}
\end{eqnarray}
In terms of the mixing angles, the couplings are
\begin{align}
&   V_{Tb}^L = s_L^u c_L^d e^{-i \phi_u} - c_L^u s_L^d e^{- i \phi_d} \,,
&& V_{Tb}^R = -c_R^u s_R^d e^{- i \phi_d} \,, \notag \\
&   V_{tB}^L = c_L^u s_L^d e^{i \phi_d} - s_L^u c_L^d e^{i \phi_u} \,, 
&& V_{tB}^R = -s_R^u c_R^d e^{i \phi_u} \,, \notag \\
&   X_{tT}^L = 0 \,,
&& X_{tT}^R = -s_R^u c_R^u e^{i \phi_u} \,, \notag \\
&   X_{bB}^L = 0 \,,
&& X_{bB}^R = -s_R^d c_R^d e^{i \phi_d} \,.
\end{align}
Under the assumption of perfect alignment $\beta - \alpha = \pi/2$, the interactions with the lightest neutral scalar $h^0$ also have the same form as in the minimal models,
\begin{eqnarray}
\mathcal{L}_{h^0} & = & -\frac{g m_T}{2M_W} \bar t \left( Y_{tT}^L P_L +  Y_{tT}^R P_R \right) T h^0 
-\frac{g m_B}{2M_W} \bar b \left( Y_{bB}^L P_L +  Y_{bB}^R P_R \right) B h^0 +\text{H.c.} \,, 
\label{ec:hlH}
\end{eqnarray}
with the couplings
\begin{align}
&   Y_{tT}^L = s_R^u c_R^u e^{i \phi_u} \,,
&& Y_{tT}^R = \frac{m_t}{m_T} s_R^u c_R^u e^{i \phi_u} \,, \notag \\
&   Y_{bB}^L = s_R^d c_R^d e^{i \phi_d} \,,
&& Y_{bB}^R = \frac{m_b}{m_B} s_R^d c_R^d e^{i \phi_d} \,.
\end{align}
The interactions with $H_1^0$ and $P_1^0$ differ by $\tan \beta$ and $-1$, $\pm i$ phase factors,
\begin{eqnarray}
\mathcal{L}_{H_1^0} & = & \frac{g m_T}{2M_W} \cot \beta \; \bar t \left( Y_{tT}^L P_L +  Y_{tT}^R P_R \right) T H_1^0 \notag \\
& & -\frac{g m_B}{2M_W} \tan \beta \; \bar b \left( Y_{bB}^L P_L +  Y_{bB}^R P_R \right) B H_1^0 +\text{H.c.} \,, \notag \\[1mm]
\mathcal{L}_{P_1^0} & = & - i \frac{g m_T}{2M_W} \cot \beta \; \bar t \left( Y_{tT}^L P_L - Y_{tT}^R P_R \right) T P_1^0 \notag \\
& & -i \frac{g m_B}{2M_W} \tan \beta \; \bar b \left( Y_{bB}^L P_L -  Y_{bB}^R P_R \right) B P_1^0 +\text{H.c.} 
\label{ec:HPlH}
\end{eqnarray}
Finally, the interactions with the charged scalar can be written as
\begin{eqnarray}
\mathcal{L}_{H^+} & = & - \frac{g m_T}{\sqrt 2 M_W} \bar T \left( \cot \beta \, Z_{Tb}^L P_L + \tan \beta \, Z_{Tb}^R P_R \right) b H^+ \notag \\[1mm]
& & -\frac{g m_B}{\sqrt 2 M_W} \bar t \left( \cot \beta \, Z_{tB}^L P_L +  \tan \beta \, Z_{tB}^R P_R \right) B H^+ +\text{H.c.} \,,
\end{eqnarray}
with the new couplings
\begin{align}
&   Z_{Tb}^L = s_L^u c_L^d e^{-i \phi_u} + \frac{s_L^d}{c_L^u} \left(s_L^{u\,2} - s_R^{u\,2} \right) e^{-i \phi_d}  \,, \notag \\
& Z_{Tb}^R = \frac{m_b}{m_T} \left[ s_L^u c_L^d e^{-i \phi_u} + \frac{c_L^u}{s_L^d} \left(s_R^{d\,2} - s_L^{d\,2} \right) e^{-i \phi_d} \right] \,, \notag \\
&   Z_{tB}^L =   \frac{m_t}{m_B} \left[ c_L^u s_L^d e^{i \phi_d} + \frac{c_L^d}{s_L^u}  \left(s_R^{u\,2} - s_L^{u\,2} \right) e^{i \phi_u}  \right] \,, \notag \\
& Z_{tB}^R = c_L^u s_L^d e^{i \phi_d} + \frac{s_L^u}{c_L^d}  \left(s_L^{d\,2} - s_R^{d\,2} \right) e^{i \phi_u}
 \,.
\end{align}


\subsection{Light-light interactions}

The inclusion of the new quarks modifies the gauge boson interactions of the third generation, leading to constraints on the mixing~\cite{Dawson:2012di,Aguilar-Saavedra:2013qpa}. These are written as
\begin{eqnarray}
\mathcal{L}_W & = & -\frac{g}{\sqrt 2} \bar t \gm \left( V_{tb}^L P_L + V_{tb}^R P_R \right) b \Wm^+ +\text{H.c.} \,, \notag \\[1mm]
\mathcal{L}_Z & = & -\frac{g}{2 c_W} \left[ \bar t \gm \left( X_{tt}^L P_L + X_{tt}^R P_R - 2 Q_t s_W^2 \right) t \right. \notag \\
& & \left. - \bar b \gm \left( X_{bb}^L P_L + X_{bb}^R P_R + 2 Q_b s_W^2 \right) b \right] \Zm \,,
\end{eqnarray}
with 
\begin{align}
&   V_{tb}^L = c_L^u c_L^d + s_L^u s_L^d e^{i (\phi_u - \phi_d)} \,,
&& V_{tb}^R = s_R^u s_R^d e^{i (\phi_u - \phi_d)} \,, \notag \\
&   X_{tt}^L = 1 \,,
&& X_{tt}^R = s_R^{u \,2} \,, \notag \\
&   X_{bb}^L = 1 \,,
&& X_{bb}^R = s_R^{d \,2} \,.
\end{align}
For small mixing angles, these couplings are close to the SM predictions. The couplings to the SM-like scalar are the same as in the minimal vector-like extensions,
\begin{eqnarray}
\mathcal{L}_{h^0} & = & -\frac{g m_t}{2M_W} Y_{tt} \; \bar t \, t \, h^0 
-\frac{g m_b}{2M_W} Y_{bb} \; \bar b \, b \, h^0 \,,
\label{ec:hll}
\end{eqnarray}
with
\begin{align}
&   Y_{tt} = c_R^{u\,2} \,,
&& Y_{bb} = c_R^{d\,2} \,.
\end{align}
$Y_{bb}$ is very close to unity due to experimental constraints, while $Y_{tt}$ can deviate from unity at the few percent level. Still, the deviations in the top quark amplitudes for $gg \to h^0$ and $h^0 \to \gamma \gamma$ caused by this difference are compensated by the contribution of the new $T$ quark, yielding a sum very close to the SM amplitude~\cite{Aguilar-Saavedra:2013qpa}.

The couplings to $H_1^0$ and $P_1^0$ are similar, but with extra $\tan \beta$, $-1$, and $\pm i$ factors, and a $\gamma_5$ matrix for the pseudo-scalar,
\begin{eqnarray}
\mathcal{L}_{H_1^0} & = & \frac{g m_t}{2M_W} Y_{tt} \cot \beta \; \bar t \, t \, H_1^0
 -\frac{g m_b}{2M_W} Y_{bb} \tan \beta \; \bar b \, b \, H_1^0 \,, \notag \\[1mm]
\mathcal{L}_{P_1^0} & = & i \frac{g m_t}{2M_W} Y_{tt} \cot \beta \; \bar t \gamma_5 t \, P_1^0
 + i \frac{g m_b}{2M_W} Y_{bb} \tan \beta \; \bar b \gamma_5 b \, P_1^0 \,.
\label{ec:HPll}
\end{eqnarray}
The interaction with the charged scalar is
\begin{eqnarray}
\mathcal{L}_{H^+} & = & - \frac{g m_t}{\sqrt 2 M_W} \bar t \left( \cot \beta \, Z_{tb}^L P_L + \tan \beta \, Z_{tb}^R P_R \right) b H^+ +\text{H.c.} \,,
\end{eqnarray}
with
\begin{align}
&   Z_{tb}^L = c_L^u c_L^d + \frac{s_L^d}{s_L^u} \left(s_L^{u\,2} - s_R^{u\,2} \right) e^{i (\phi_u - \phi_d)}  \,,
&& Z_{tb}^R = \frac{m_b}{m_t} \left[ c_L^u c_L^d + \frac{s_L^u}{s_L^d} \left(s_L^{d\,2} - s_R^{d\,2} \right) e^{i (\phi_u - \phi_d)} \right] \,. \notag \\
\end{align}
For masses larger than $2 m_t$, the heavy scalars $H_1^0$ and $P_1^0$ will dominantly decay into $t \bar t$ or $b \bar b$, depending on $\tan \beta$. (For lighter $H_1^0$, the decay into $h^0 h^0$ may be sizeable~\cite{Djouadi:2005gj}.)  The charged scalar is expected to decay mainly into $t \bar b$.


\subsection{Heavy-heavy interactions}

The couplings between the two heavy quarks are not involved in production nor decay processes; we collect here only for completeness. The Lagrangians have the same form as for the light-light interactions but replacing $t$ by $T$ and $b$ by $B$. The corresponding couplings are
\begin{align}
&   V_{TB}^L = c_L^u c_L^d + s_L^u s_L^d e^{-i (\phi_u - \phi_d)} \,,
&& V_{TB}^R = c_R^u c_R^d  \,, \notag \\
&   X_{TT}^L = 1 \,,
&& X_{TT}^R = c_R^{u \,2} \,, \notag \\
&   X_{BB}^L = 1 \,,
&& X_{BB}^R = c_R^{d \,2} \,, \notag \\
&   Y_{TT} = s_R^{u\,2} \,,
&& Y_{BB} = s_R^{d\,2} \,, \notag \\
&   Z_{TB}^L = s_L^u s_L^d e^{-i (\phi_u - \phi_d)} + \frac{c_L^d}{c_L^u} \left(s_R^{u\,2} - s_L^{u\,2} \right)   \,, 
&& Z_{TB}^R = \frac{m_B}{m_T} \left[ s_L^u s_L^d e^{-i (\phi_u - \phi_d)} + \frac{c_L^u}{c_L^d} \left(s_R^{d\,2} - s_L^{d\,2} \right) \right] \,.
\end{align}


\section{Interactions for two scalar doublets plus a singlet}
\label{sec:3}

We consider here the mixing with one additional scalar singlet 
$\widetilde \nu_R$. We do not take the most general $3 \times 3$ unitary transformations for the scalars and pseudo-scalars, but instead we use simple two-angle rotations that ensure that there is a SM-like Higgs boson $h^0$ and still yield a richer phenomenology than in the two-scalar doublet model. We write
\begin{eqnarray}
H_u^0 & = & \frac{1}{\sqrt 2} \left[ \cos \alpha \, h^0 + \sin \alpha \cos \theta' \, H_1^0 + \sin \alpha \sin \theta' \, H_2^0 \right. \notag \\
& & \left. + i \sin \beta \, G^0 + i \cos \beta \cos \theta \, P_1^0 + i \cos \beta \sin \theta \, P_2^0 \right] \,, \notag \\
H_d^0 & = & \frac{1}{\sqrt 2} \left[ -\sin\alpha \, h^0 + \cos \alpha \cos \theta' \, H_1^0 + \cos \alpha \sin \theta' \, H_2^0 \right. \notag \\
& & \left.  - i \cos \beta \, G^0 + i \sin \beta \cos \theta \, P_1^0 + i \sin \beta \sin \theta \, P_2^0 \right] \,, \notag \\
\widetilde \nu_R & = &  \frac{1}{\sqrt 2} \left[ - \sin \theta' H_1^0 + \cos \theta' H_2^0 - i \sin \theta P_1^0 + i \cos \theta P_2^0 \right] \,.
\end{eqnarray}
The interactions with $h^0$ are unchanged with respect to the previous section. The light-heavy interactions for up-type quarks with the scalars and pseudo-scalars can be obtained from the interactions with $H_1^0$ and $P_1^0$ in eq. (\ref{ec:HPlH}), respectively, with the replacements
\begin{align}
- \cot \beta & \to - \cot \beta \cos \theta' + \frac{1}{\kappa_R} \sin \theta'  && (H_1^0) \,, \notag \\
 - \cot \beta & \to - \cot \beta \sin \theta' - \frac{1}{\kappa_R} \cos \theta'  && (H_2^0) \,, \notag \\
  \cot \beta & \to  \cot \beta \cos \theta + \frac{1}{\kappa_R} \sin \theta  && (P_1^0) \,, \notag \\
  \cot \beta & \to  \cot \beta \sin \theta - \frac{1}{\kappa_R} \cos \theta  && (P_2^0) \,,
\label{ec:repl1}
\end{align}
where we have defined $\kappa_R = v_R / v$.
 The  terms proportional to $1/\kappa_R$ arise from the vector-like doublet coupling to the scalar singlet in eq.~(\ref{ec:lY2}), which generates the $(4,4)$ entries in the mass matrices that approximately equal the heavy quark masses. Analogously, for the down-type quarks the interactions are obtained by replacing
\begin{align}
\tan \beta & \to \tan \beta \cos \theta' + \frac{1}{\kappa_R} \sin \theta'  & (H_1^0) \,, \notag \\
\tan \beta & \to \tan \beta \sin \theta' - \frac{1}{\kappa_R} \cos \theta'  &  (H_2^0) \,, \notag \\
\tan \beta & \to  \tan \beta \cos \theta + \frac{1}{\kappa_R} \sin \theta  &  (P_1^0) \,, \notag \\
\tan \beta & \to  \tan \beta \sin \theta - \frac{1}{\kappa_R} \cos \theta  &  (P_2^0) \,,
\label{ec:repl2}
\end{align}
in the scalar and pseudo-scalar interactions written in eq. (\ref{ec:HPlH}).

The light-light interactions are slightly more involved. For the top quark, they are obtained from eq.~(\ref{ec:HPll}) by replacing
\begin{align}
- \cot \beta \, Y_{tt} & \to - \cot \beta \cos \theta' \, Y_{tt} - \frac{1}{\kappa_R} \sin \theta' \, (1-Y_{tt})  && (H_1^0) \,, \notag \\
 - \cot \beta \, Y_{tt} & \to - \cot \beta \sin \theta' \, Y_{tt} + \frac{1}{\kappa_R} \cos \theta' \, (1-Y_{tt})  && (H_2^0) \,, \notag \\
  \cot \beta \, Y_{tt} & \to  \cot \beta \cos \theta\, Y_{tt} - \frac{1}{\kappa_R} \sin \theta \, (1-Y_{tt})  && (P_1^0) \,, \notag \\
  \cot \beta \, Y_{tt} & \to  \cot \beta \sin \theta\, Y_{tt} + \frac{1}{\kappa_R} \cos \theta  \, (1-Y_{tt}) && (P_2^0) \,,
 \label{ec:repl3}
\end{align}
and for the bottom quark,
\begin{align}
\tan \beta \, Y_{bb} & \to \tan \beta \cos \theta' \, Y_{bb} - \frac{1}{\kappa_R} \sin \theta' \, (1-Y_{bb}) & (H_1^0) \,, \notag \\
\tan \beta \, Y_{bb} & \to \tan \beta \sin \theta' \, Y_{bb} + \frac{1}{\kappa_R} \cos \theta'  \, (1-Y_{bb})&  (H_2^0) \,, \notag \\
\tan \beta \, Y_{bb} & \to  \tan \beta \cos \theta \, Y_{bb} - \frac{1}{\kappa_R} \sin \theta \, (1-Y_{bb}) &  (P_1^0) \,, \notag \\
\tan \beta \, Y_{bb} & \to  \tan \beta \sin \theta \, Y_{bb} + \frac{1}{\kappa_R} \cos \theta \, (1-Y_{bb}) &  (P_2^0) \,.
\label{ec:repl4}
\end{align}
In addition to the decays into $t \bar t$ and $b \bar b$, mediated by the couplings in (\ref{ec:repl3}) and (\ref{ec:repl4}), the scalars $H_k^0$, $k=1,2$, can have more exotic decay modes such as $h^0 h^0$~\cite{Fidalgo:2011ky}.

Finally, the heavy-heavy interactions have the same form as light-light interactions but replacing the quark masses and $Y_{tt} \to Y_{TT}$, $Y_{bb} \to Y_{BB}$.


\section{Decay of the heavy quarks}
\label{sec:4}

The heavy quarks $T$ and $B$ can decay into SM gauge or Higgs bosons plus a lighter quark, cf.~(\ref{ec:stddec}), as in the minimal models with a single Higgs doublet. Provided the channels are kinematically allowed, they can also decay into the extra scalars plus a top or bottom quark, 
\begin{align}
& T \to H_k^0 t \,, \quad T \to P_k^0 t \,,\quad T \to H^+ b\,, \notag \\
& B \to H_k^0 b \,, \quad B \to P_k^0 b  \,, \quad B \to H^- t   \,,
\end{align}
with $k=1,2$.
The expressions for the partial widths are collected in appendix~\ref{sec:a}.
They depend on the mixing angles $s_R^u$ and $s_R^d$, the mixing in the scalar sector and the heavy quark and (pseudo-)scalar masses. If $T$ and $B$ are much heavier than $H_k^0$, $P_k^0$ and $H^\pm$, the dependence on the masses is mild. We will therefore fix the quark masses to $m_T = m_B = 1$ TeV,\footnote{The quark mixing induces a small splitting between the $T$ and $B$ masses~\cite{Aguilar-Saavedra:2013qpa}, which plays no role here and is ignored for simplicity.} and new scalar masses to $M_{H_k^0} = M_{P_k^0} = M_{H^\pm} = 0.5$ TeV.

The angle $\theta_R^d$ determines the size of the charged current mixing of the $T$ quark and the neutral current mixing of the $B$ quark. Conversely, the angle $\theta_R^u$ determines the charged current mixing of the $B$ quark and neutral current mixing of the $T$ quark. Therefore, the decays of either $T$ or $B$ depend on both mixing angles. We will use several representative benchmarks for the quark mixing, all of them with the phases $\phi_u$ and $\phi_d$ set to zero:
\begin{itemize}
\item[(i)] Equal mixing $s_R^u = s_R^d$. We take both of them equal to 0.05, fulfilling indirect constraints~\cite{Aguilar-Saavedra:2013qpa,Chen:2017hak}. The couplings are collected in table~\ref{tab:coup-eq}.
\item[(ii)] Dominant mixing in the up sector, as it is expected from the quark mass hierarchy. We take $s_R^u = 0.05$, $s_R^d = 0.01$ as well as the limit case $s_R^u = 0.05$, $s_R^d \sim 0$.  The couplings are collected in table~\ref{tab:coup-up}.
\item[(iii)] Dominant mixing in the down sector. This inverted hierarchy needs some fine tuning of parameters, but is studied for completeness. We take $s_R^d = 0.05$, $s_R^u =0.01$ and the limit case $s_R^d = 0.05$, $s_R^u \sim 0$, giving the couplings in table~\ref{tab:coup-down}.
\end{itemize}
\begin{table}[t]
\begin{center}
\begin{tabular}{ccccccc}
                & $L$          & $R$ & \quad & & $L$          & $R$    \\
$V_{Tb}$ & $0.0084$ & $-0.05$     && $V_{tB}$  & $-0.0084$ & $-0.05$   \\
$X_{tT}$  &     0          & $-0.05$     && $X_{bB}$ & 0               & $-0.05$    \\
$Y_{tT}$  & $0.05$     & $0.0086$  && $Y_{bB}$ & $0.05$      & $0.00024$   \\
$Z_{Tb}$ & $0.0086$ & $0.05$      && $Z_{tB}$  &  $0.048$   & $0.00022$ 
\end{tabular}
\caption{Couplings for the equal-mixing scenario with $s_R^u = 0.05$, $s_R^d = 0.05$ (implying $s_L^u  =  0.0086$, $s_L^d  =  0.00024$ for $m_{T,B} = 1$ TeV).}
\label{tab:coup-eq}
\end{center}
\end{table}
\begin{table}[htb]
\begin{center}
\begin{tabular}{ccccccc}
\multicolumn{7}{c}{dominant up mixing} \\
                & $L$          & $R$ & \quad & & $L$          & $R$    \\
$V_{Tb}$ & $0.0086$ & $-0.01$     && $V_{tB}$  & $-0.0086$ & $-0.05$   \\
$X_{tT}$  &     0          & $-0.05$       && $X_{bB}$ & 0               & $-0.01$    \\
$Y_{tT}$  & $0.05$     & $0.0086$    && $Y_{bB}$ & $0.01$      & $5 \times 10^{-5}$   \\
$Z_{Tb}$ & $0.0086$ & $0.01$   && $Z_{tB}$  &  $0.048$   & $5 \times 10^{-5}$ \\[3mm]
\multicolumn{7}{c}{only up mixing} \\
                & $L$          & $R$ & \quad & & $L$          & $R$    \\
$V_{Tb}$ & $0.0086$ & $\sim 0$     && $V_{tB}$  & $-0.0086$ & $-0.05$   \\
$X_{tT}$  &     0          & $-0.05$       && $X_{bB}$ & 0               & $\sim 0$    \\
$Y_{tT}$  & $0.05$     & $0.0086$    && $Y_{bB}$ & $\sim 0$      & $\sim 0$   \\
$Z_{Tb}$ & $0.0086$ & $0.00004$   && $Z_{tB}$  &  $0.048$   & $\sim 0$ 
\end{tabular}
\caption{Top: Couplings for the up-mixing scenario with $s_R^u = 0.05$, $s_R^d = 0.01$ for $m_{T,B} = 1$ TeV (for which $s_L^u  =  0.0086$, $s_L^d  = 5 \times 10^{-5}$). Bottom: the same for $s_R^u = 0.05$, $s_R^d \sim 0 $ ($s_L^u  =  0.0086$, $s_L^d  \sim 0$).}
\label{tab:coup-up}
\end{center}
\end{table}
\begin{table}[htb]
\begin{center}
\begin{tabular}{ccccccc}
\multicolumn{7}{c}{dominant down mixing} \\
                & $L$          & $R$ & \quad & & $L$          & $R$    \\
$V_{Tb}$ & $0.0015$ & $-0.05$       && $V_{tB}$  & $-0.0015$ & $-0.01$   \\
$X_{tT}$  &     0          & $-0.01$       && $X_{bB}$ & 0               & $-0.05$    \\
$Y_{tT}$  & $0.01$   & $0.0017$      && $Y_{bB}$ & $0.05$      & $0.00024$   \\
$Z_{Tb}$ & $0.0017$   & $0.05$         && $Z_{tB}$  &  $0.0097$   & $0.00024$ \\
\multicolumn{7}{c}{only down mixing} \\
                & $L$          & $R$ & \quad & & $L$          & $R$    \\
$V_{Tb}$ & $-0.00024$ & $-0.05$       && $V_{tB}$  & $0.00024$ & $\sim 0$   \\
$X_{tT}$  &     0          & $\sim 0$       && $X_{bB}$ & 0               & $-0.05$    \\
$Y_{tT}$  & $\sim 0$   & $\sim 0$      && $Y_{bB}$ & $0.05$      & $0.00024$   \\
$Z_{Tb}$ & $\sim 0$   & $0.05$         && $Z_{tB}$  &  $4 \times 10^{-4}$   & $0.00024$ 
\end{tabular}
\caption{Top: Couplings for the down-mixing scenario with $s_R^u =0.01$, $s_R^d =0.05$ for $m_{T,B} = 1$ TeV (for which $s_L^u  = 0.0086$, $s_L^d  =0.00024$). Bottom: the same, for $s_R^u \sim 0$, $s_R^d =0.05$ ($s_L^u  \sim 0$, $s_L^d  =0.00024$).}
\label{tab:coup-down}
\end{center}
\end{table}
We consider in turn the simpler model with only two scalar doublets (i.e. no mixing with the singlet) and with two scalar doublets plus a singlet.


\subsection{Two scalar doublets}
\label{sec:4.1}

\begin{figure}[htb]
\begin{center}
\begin{tabular}{cc}
\includegraphics[height=5.25cm,clip=]{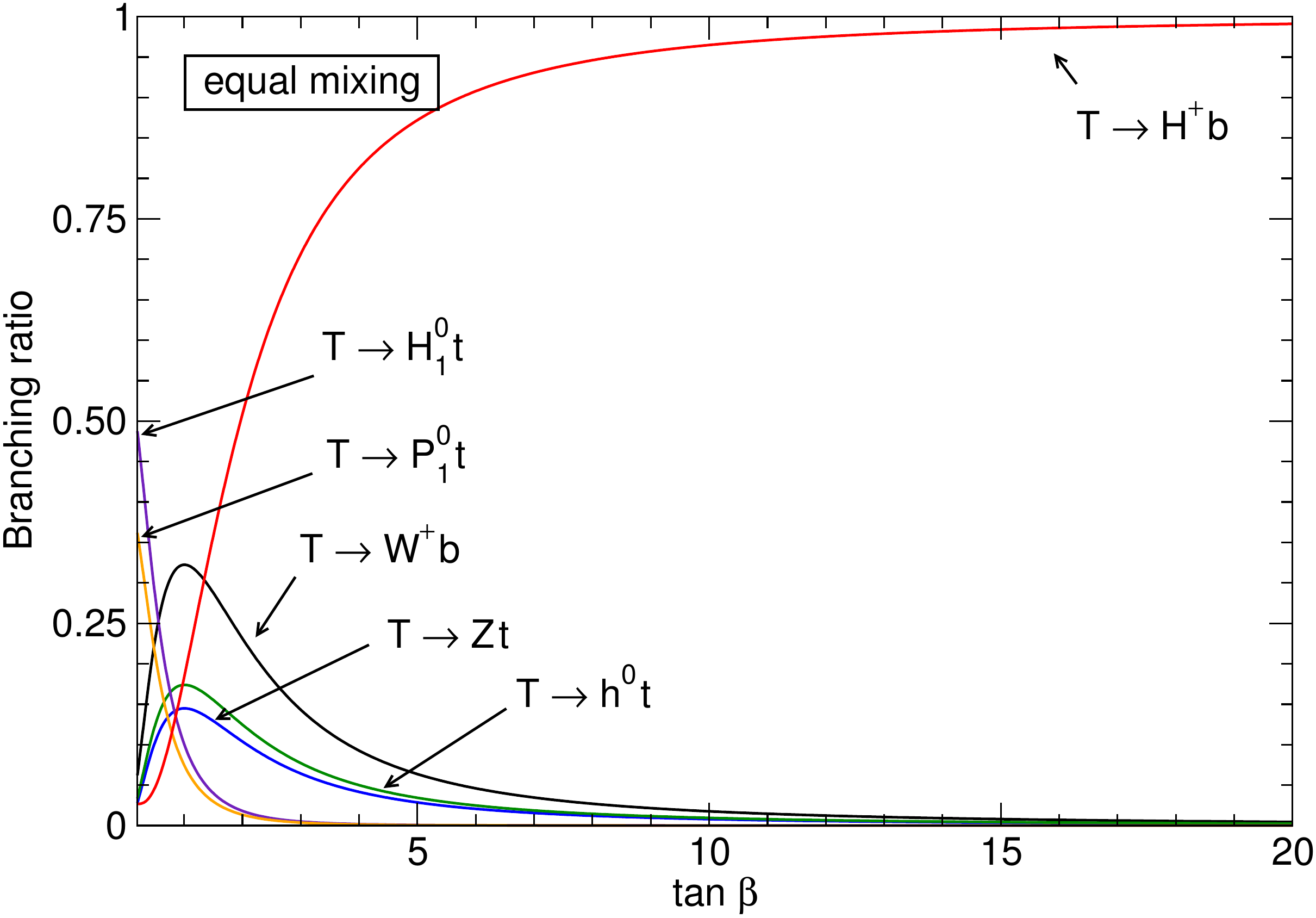} &
\includegraphics[height=5.25cm,clip=]{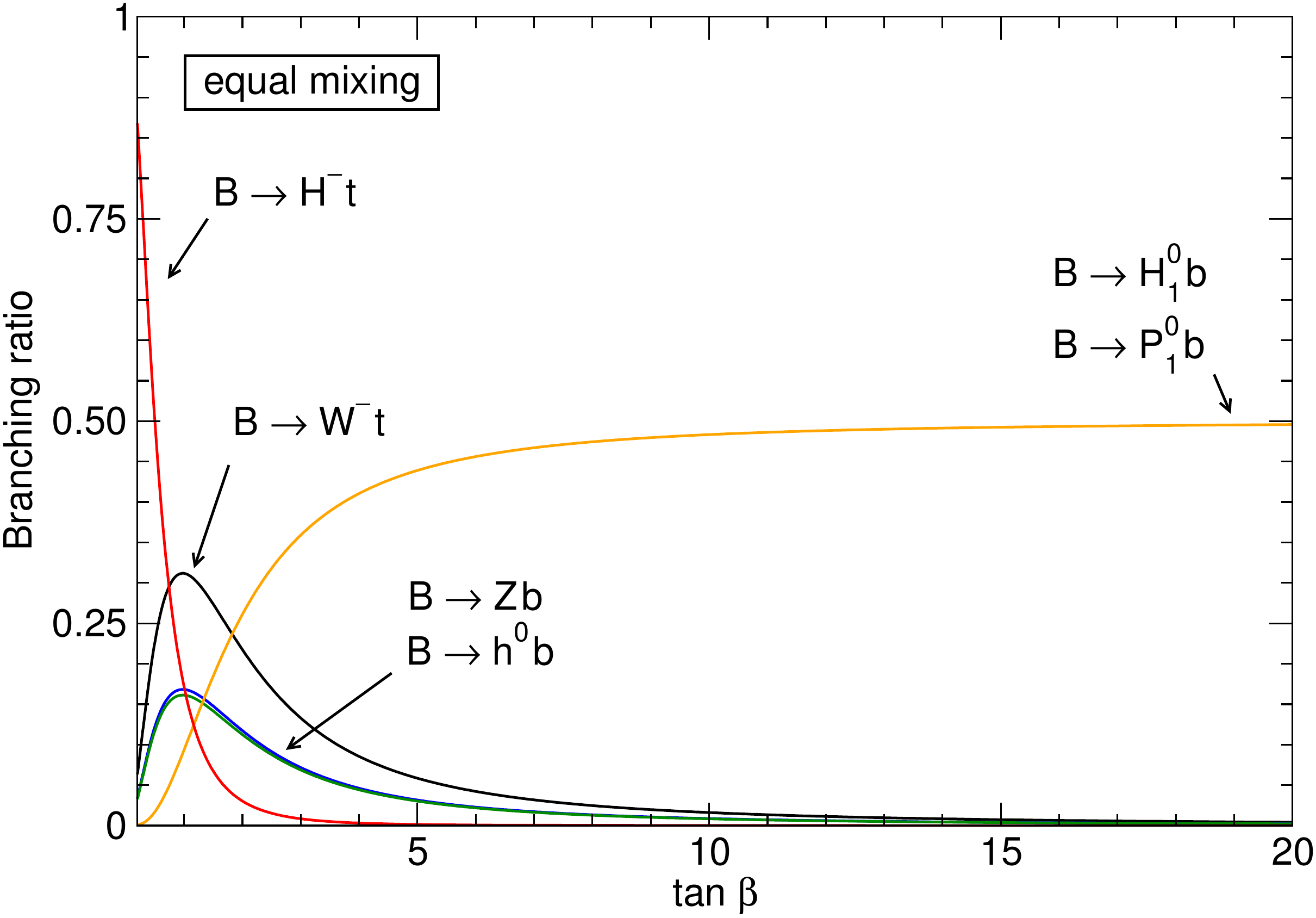} \\[2mm]
\includegraphics[height=5.25cm,clip=]{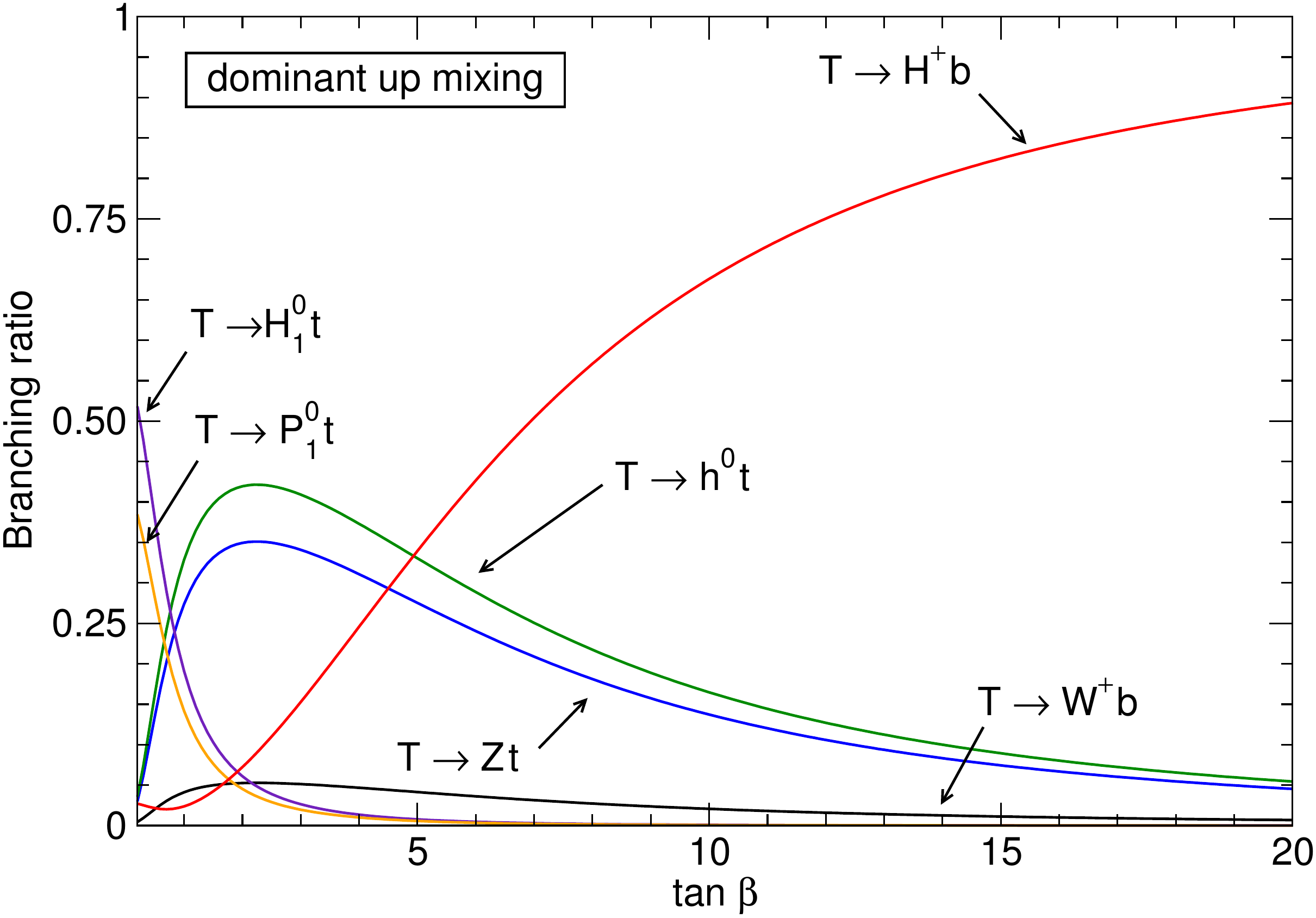} &
\includegraphics[height=5.25cm,clip=]{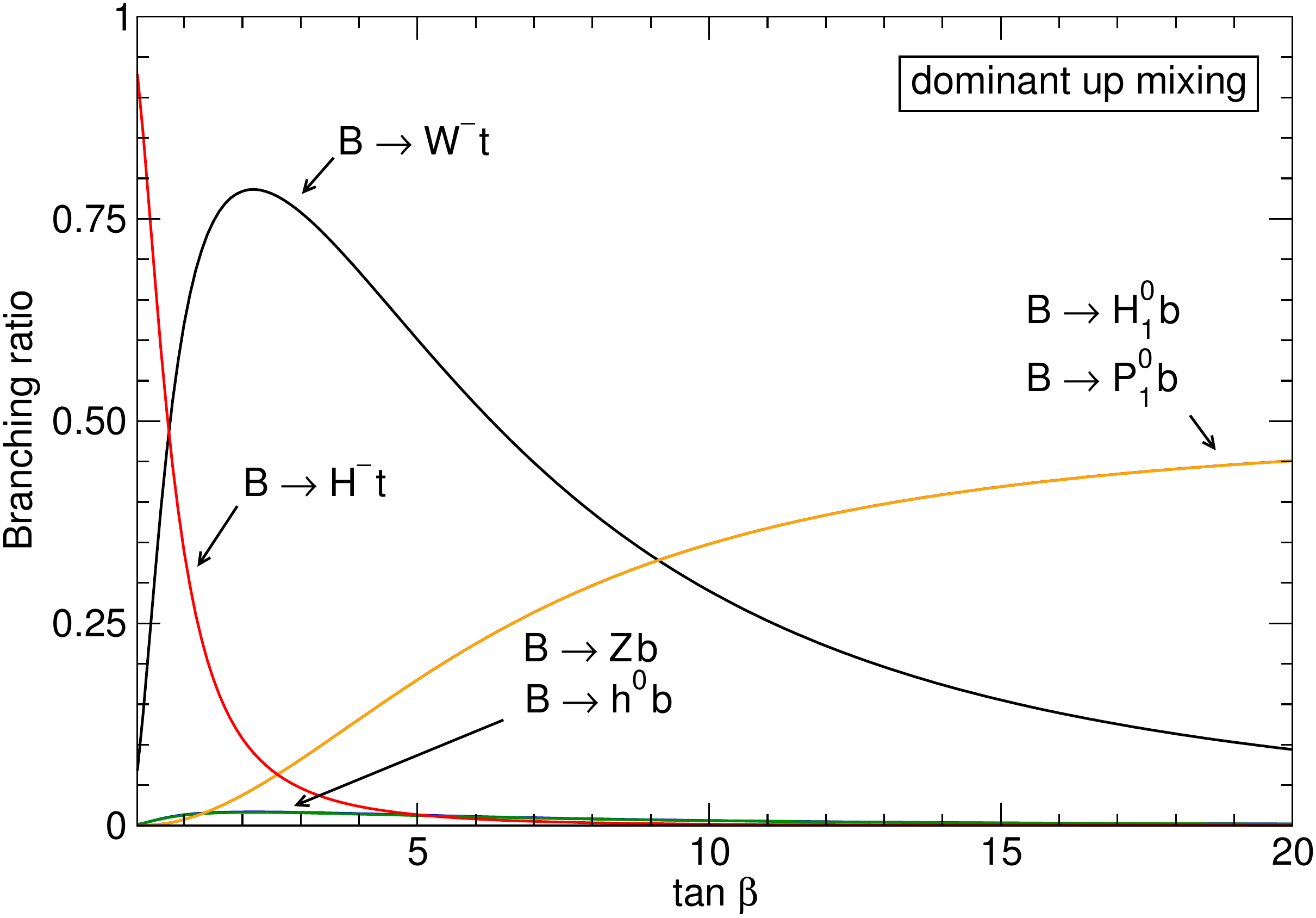} \\[2mm]
\includegraphics[height=5.25cm,clip=]{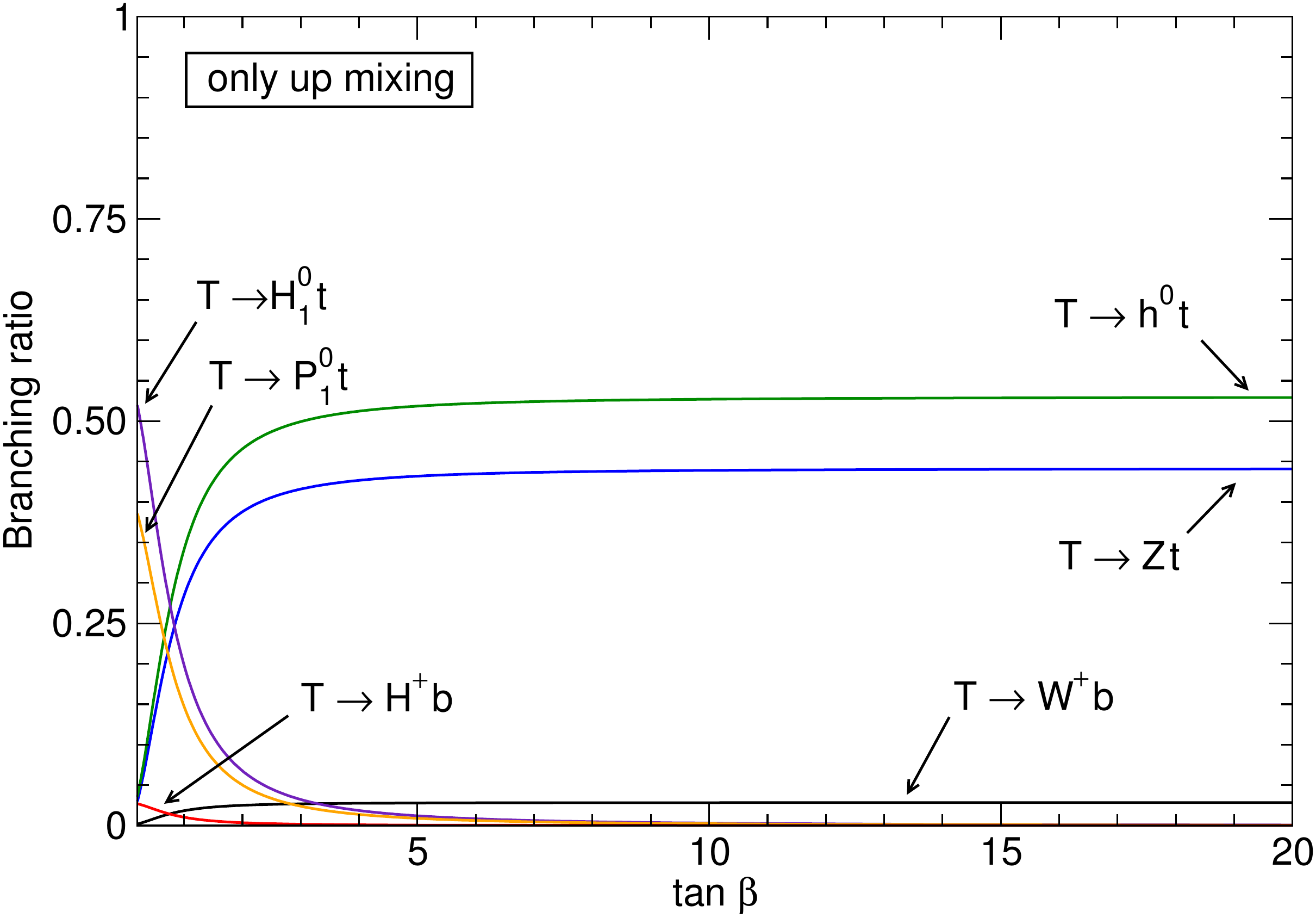} &
\includegraphics[height=5.25cm,clip=]{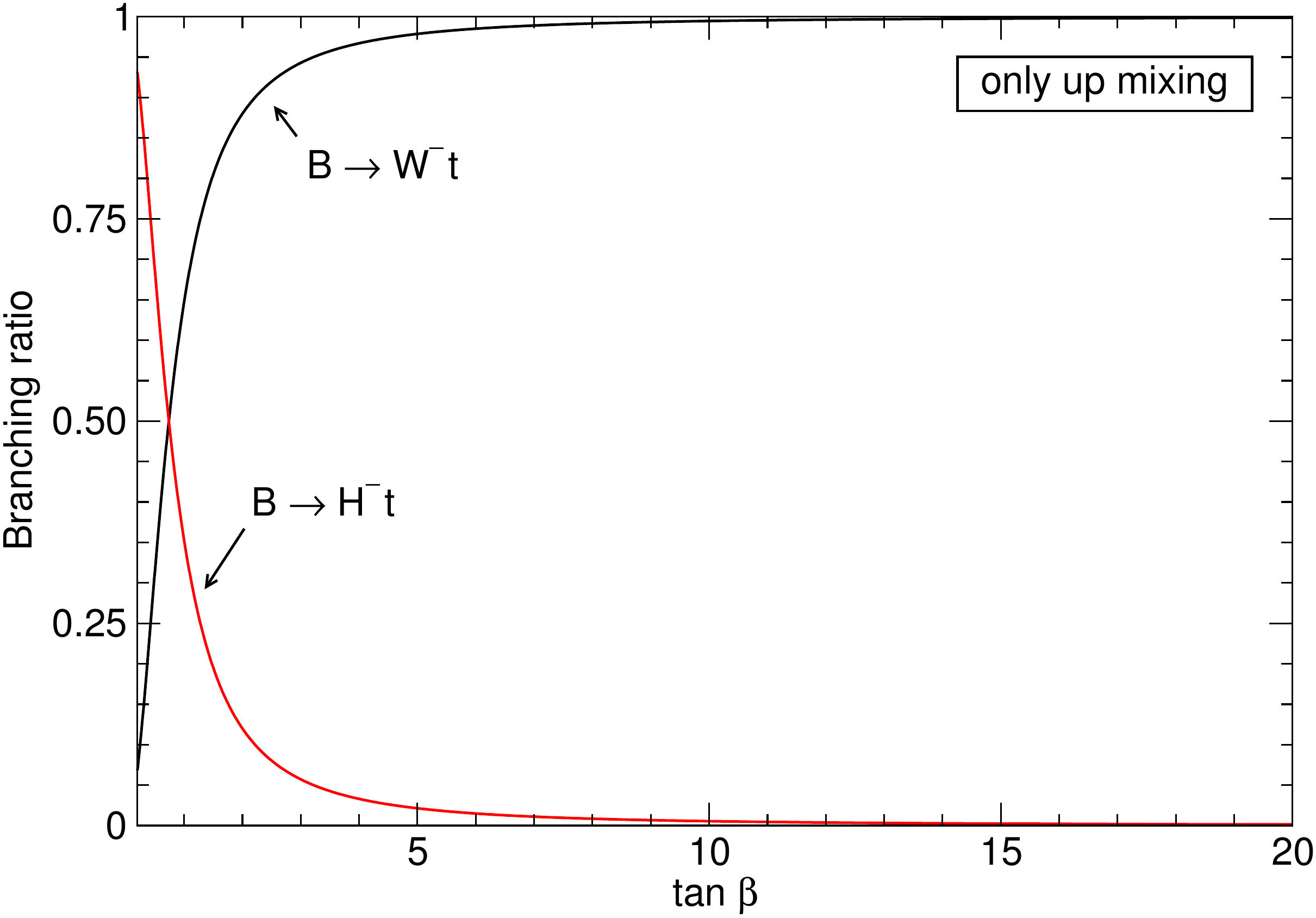} 
\end{tabular}
\caption{Dependence of the $T$ and $B$ branching ratios on $\tan \beta$, for the equal mixing and dominant up mixing scenarios of tables~\ref{tab:coup-eq} and \ref{tab:coup-up}.}
\label{fig:BR1}
\end{center}
\end{figure}

\begin{figure}[htb]
\begin{center}
\begin{tabular}{cc}
\includegraphics[height=5.25cm,clip=]{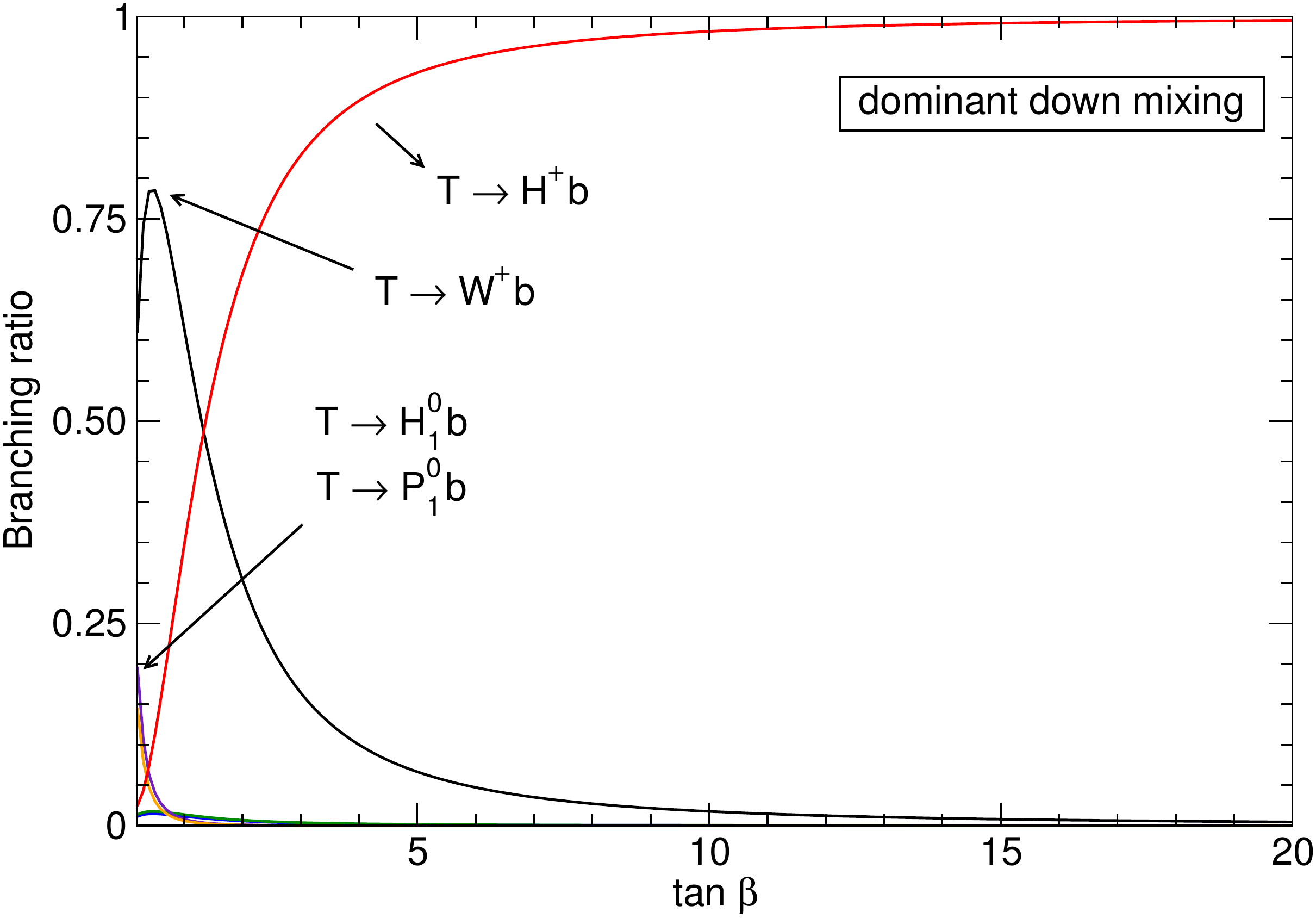} &
\includegraphics[height=5.25cm,clip=]{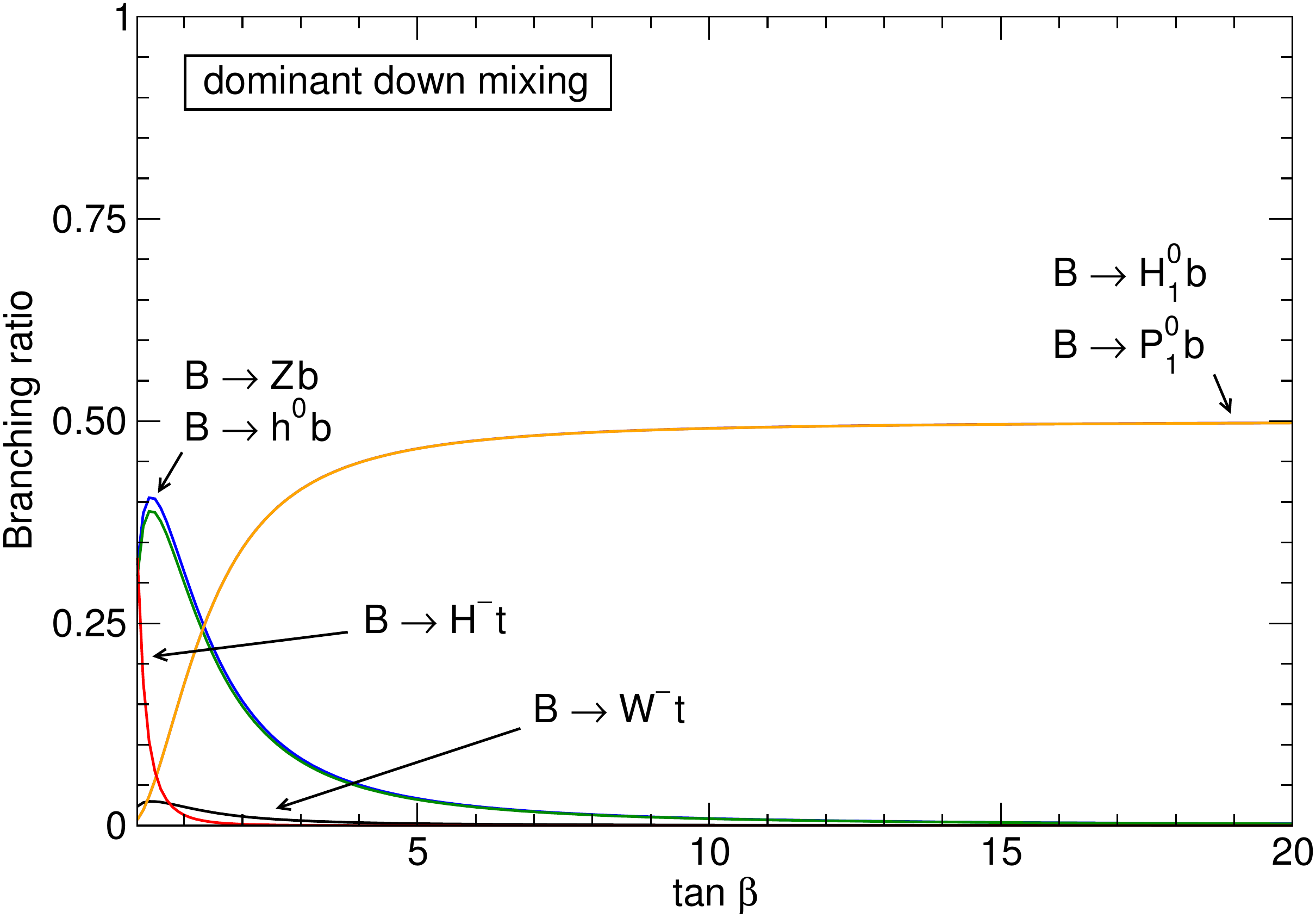} \\[2mm]
\includegraphics[height=5.25cm,clip=]{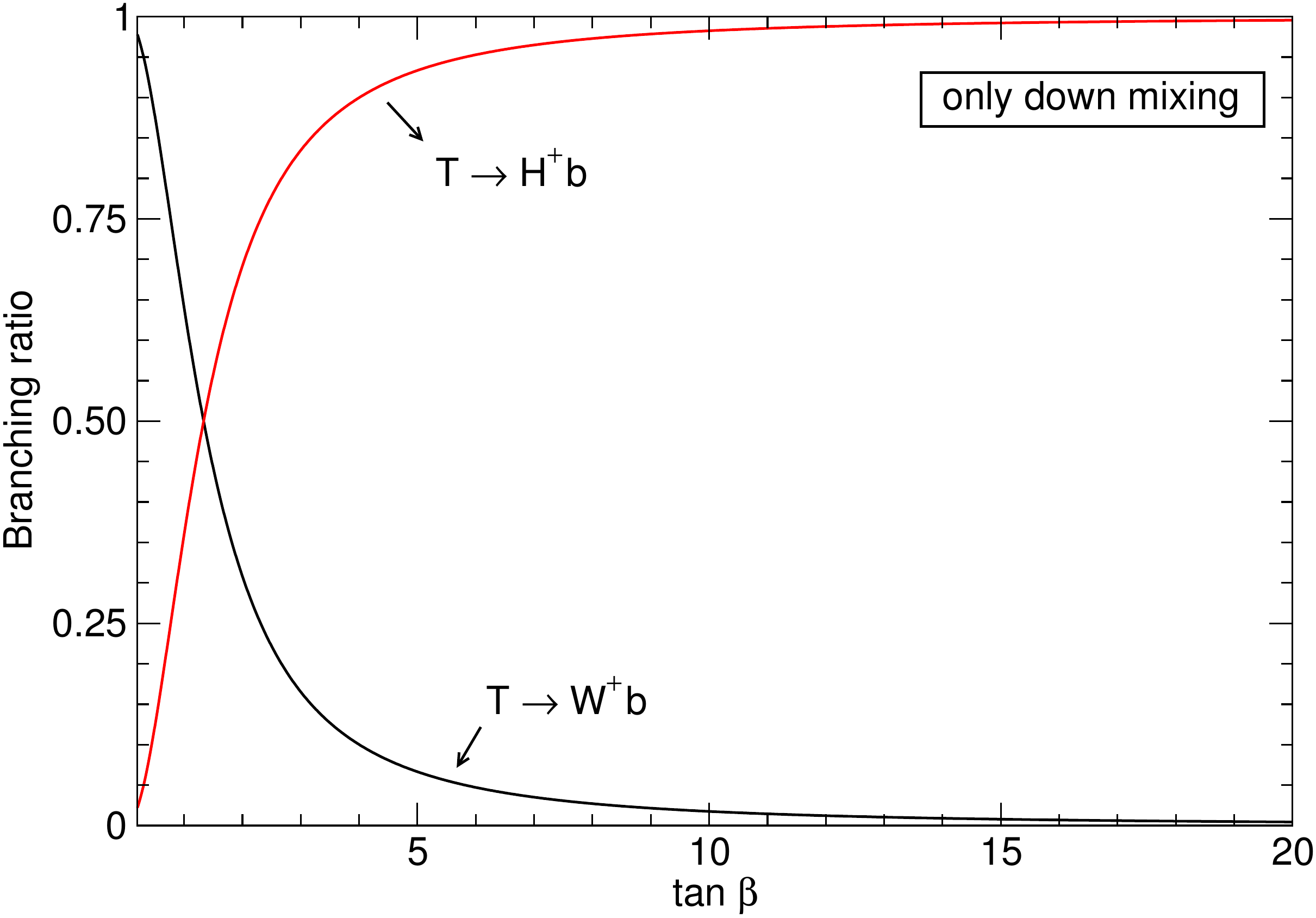} &
\includegraphics[height=5.25cm,clip=]{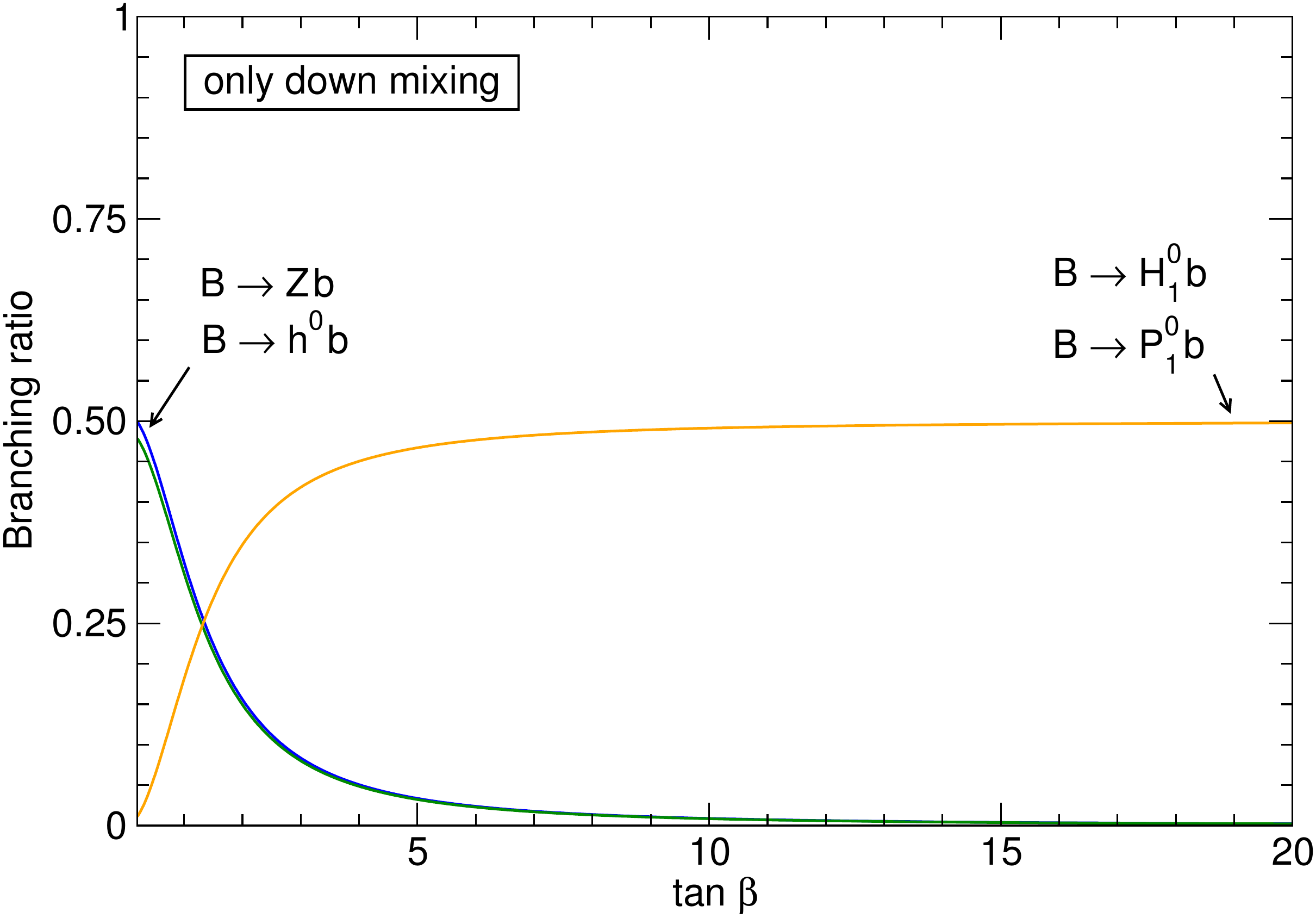} 
\end{tabular}
\caption{Dependence of the $T$ and $B$ branching ratios on $\tan \beta$, for the dominant down mixing scenarios of table~\ref{tab:coup-down}.}
\label{fig:BR2}
\end{center}
\end{figure}

For each of the quark mixing benchmarks in tables~\ref{tab:coup-eq}--\ref{tab:coup-down}, we plot in figures~\ref{fig:BR1} and \ref{fig:BR2} the dependence of the branching ratios on $\tan \beta$. 
The results can be understood from the relative size of the couplings and the $\tan \beta$ factors in the Lagrangian. In most cases it is found that the new channels with neutral scalars $H_1^0$ or $P_1^0$ and the channel with the charged one $H^\pm$ in the final states do not compete among themselves, but only with the standard ones. For $T$ decays we have:
\begin{itemize}
\item For equal mixing and $\tan \beta \sim 1$, all $T$ decay channels are open and the branching ratios are roughly of the same order.  As $\tan \beta$ gets large, the $H^+ b$ mode dominates because the $\tan \beta \, Z_{Tb}^R$ factor in the coupling, with $Z_{Tb}^R \simeq s_R^d$, gets large while the decays to $H_1^0 t$ and $P_1^0 t$ are suppressed by $\cot^2 \beta$. For small $\tan \beta$ the opposite occurs.
\item For dominant up mixing, the $V_{Tb}$ and $Z_{Tb}$ couplings are small, hence the only relevant modes are the neutral ones, and $T \to H^+ b$ at large $\tan \beta$ if $s_R^d$ is not too small. For large $\tan \beta$, the decays into $H_1^0 t$ and $P_1^0 t$ are suppressed by $\cot^2 \beta$ and are negligible, whereas for low $\tan \beta$ they dominate.
\item For dominant down mixing, the $X_{tT}$ and $Y_{tT}$ couplings are close to zero and the charged current modes dominate. The decay into $H^+ b$ is determined by the $\tan \beta \, Z_{Tb}^R$ coupling, therefore it is enhanced at large $\tan \beta$ and suppressed at low $\tan \beta$, as in the equal mixing scenario.
\end{itemize}
For $B$ decays, the situation is reversed because the dependence on $\tan \beta$ is the opposite as for the $T$ quark. We can see that:
\begin{itemize}
\item For equal mixing and $\tan \beta \sim 1$ all $B$ decay modes have branching ratios of the same order. At large $\tan \beta$ the factor $Z_{tB}^L \cot \beta$ in the coupling, with $Z_{tB}^L \sim s_R^u$, 
is suppressed (in this case the coupling $Z_{tB}^R$ is very small) while the decays to $H_1^0 b$ and $P_1^0 b$ benefit from the $\tan^2 \beta$ enhancement and therefore dominate. For small $\tan \beta$ the opposite happens, and $B \to H^- t$ dominates.
\item For dominant up mixing the neutral couplings $X_{bB}$ and $Y_{bB}$ are small, and decays into $Zb$ and $h^0 b$ are negligible. The $Z_{tB}^L \cot \beta$ coupling is suppressed at large $\tan \beta$ and makes the $B \to H^- t$ channel small; for the same reason it dominates at low $\tan \beta$. If the mixing in the down sector is not too small, at large $\tan \beta$ the decays into $H_1^0 b$ and $P_1^0 b$ can be important, otherwise $B \to W^- t$ is the leading channel at large $\tan \beta$.
\item For dominant down mixing the charged current couplings $V_{tB}$ and $Z_{tB}$ of the $B$ quark are very small, so we mainly have the neutral decays. $B \to H_1^0 b$ and $B \to P_1^0 b$ are enhanced at large $\tan \beta$ and suppressed in the small $\tan \beta$ region. 
\end{itemize}
The neutral (pseudo-)scalars $H_1^0$ and $P_1^0$ produced in the heavy quark decays are expected to decay mainly into $t \bar t$ (low $\tan \beta$, and provided the channel is kinematically open) and $b \bar b$ (high $\tan \beta$), with equal branching ratios for $\tan \beta \simeq 6$. The partial widths for the decays are given in appendix~\ref{sec:a}.


\subsection{Two scalar doublets plus a singlet}
\label{sec:4.2}

In this case there are three additional parameters: the ratio of VEVs $\kappa_R$ and the two mixing angles $\theta$ and $\theta'$. However, if we require that the Yukawa coupling $y_{44}$ of the quark doublet to the scalar singlet is of order one at most, then $\kappa_R \gtrsim 6$ for $m_{T,B} = 1$ TeV and the latter terms in eqs. (\ref{ec:repl1}) and (\ref{ec:repl2}) are small. Consequently, for the situations of interest the parameter $\kappa_R$ has little influence on the $T$ and $B$ decay branching ratios. 

\begin{figure}[t]
\begin{center}
\begin{tabular}{cc}
\includegraphics[height=5.25cm,clip=]{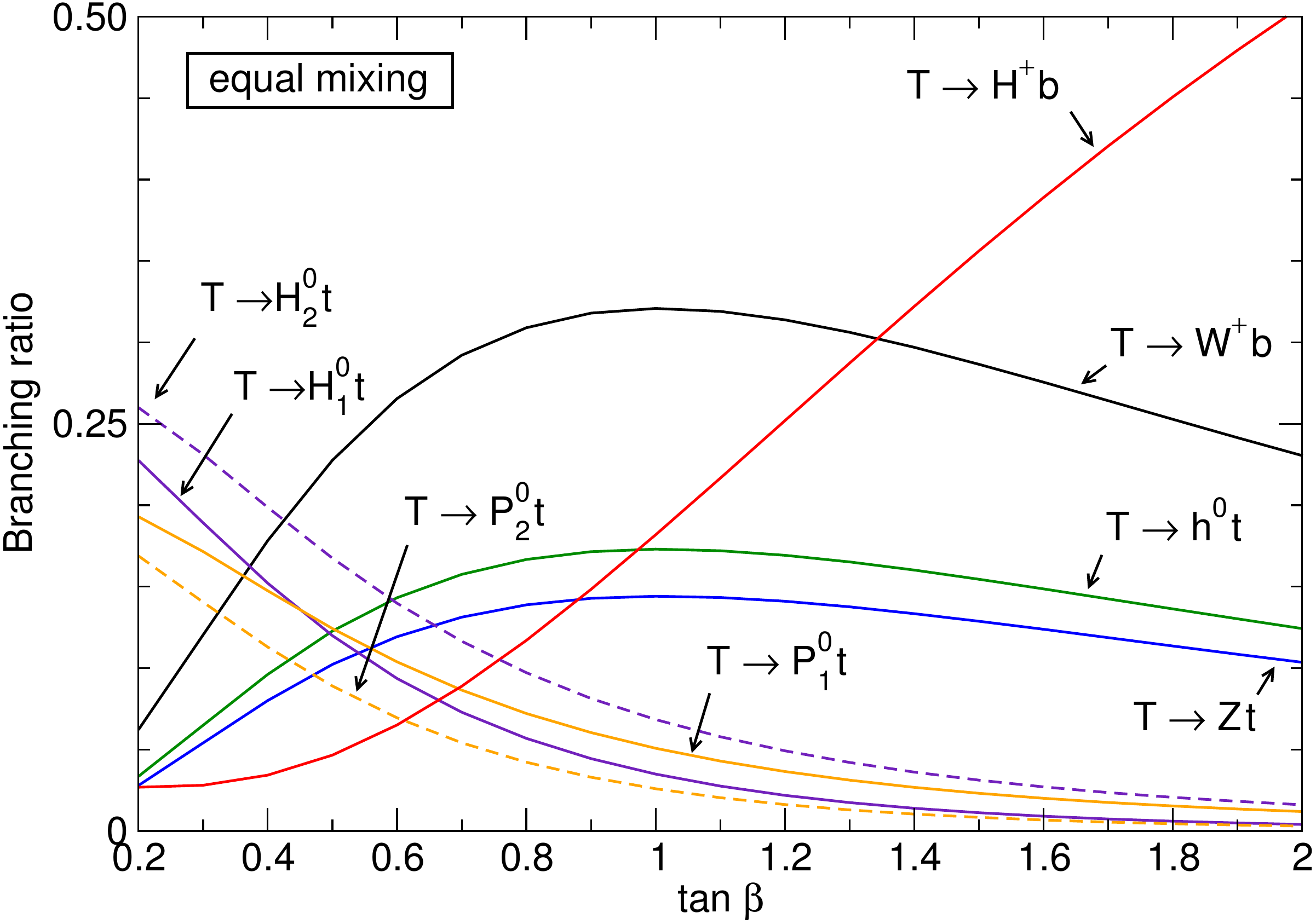} &
\includegraphics[height=5.25cm,clip=]{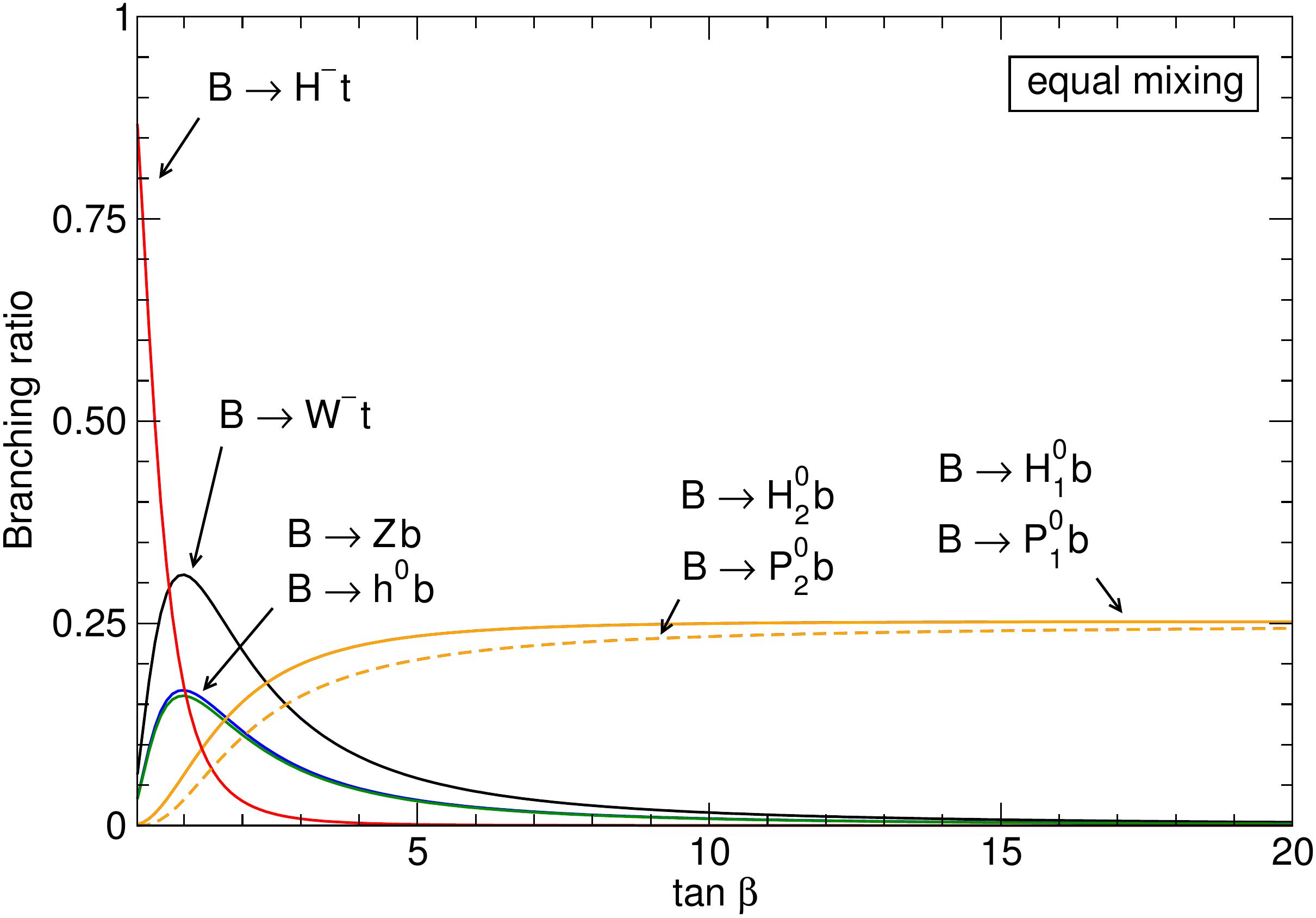} \\
\includegraphics[height=5.25cm,clip=]{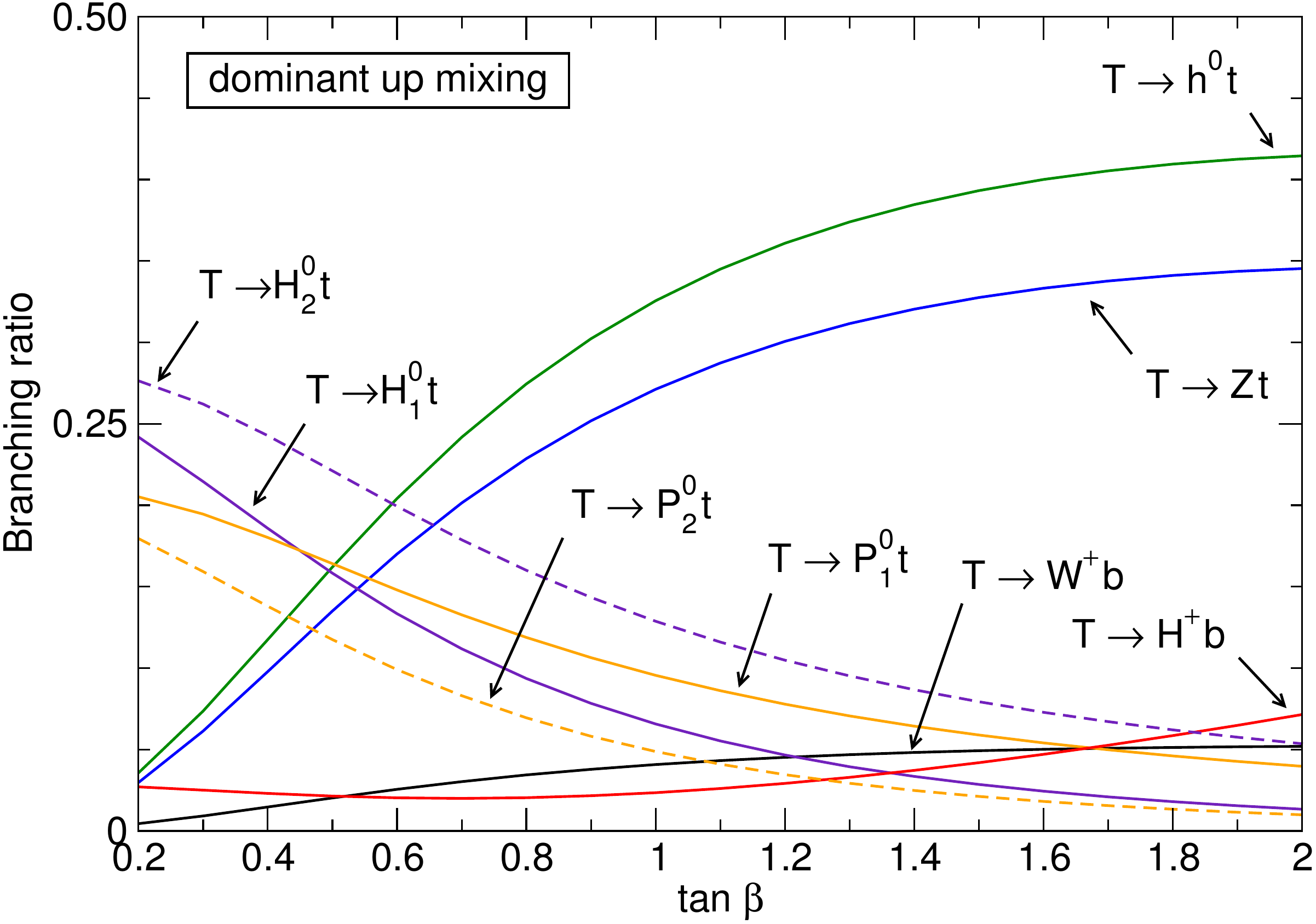} &
\includegraphics[height=5.25cm,clip=]{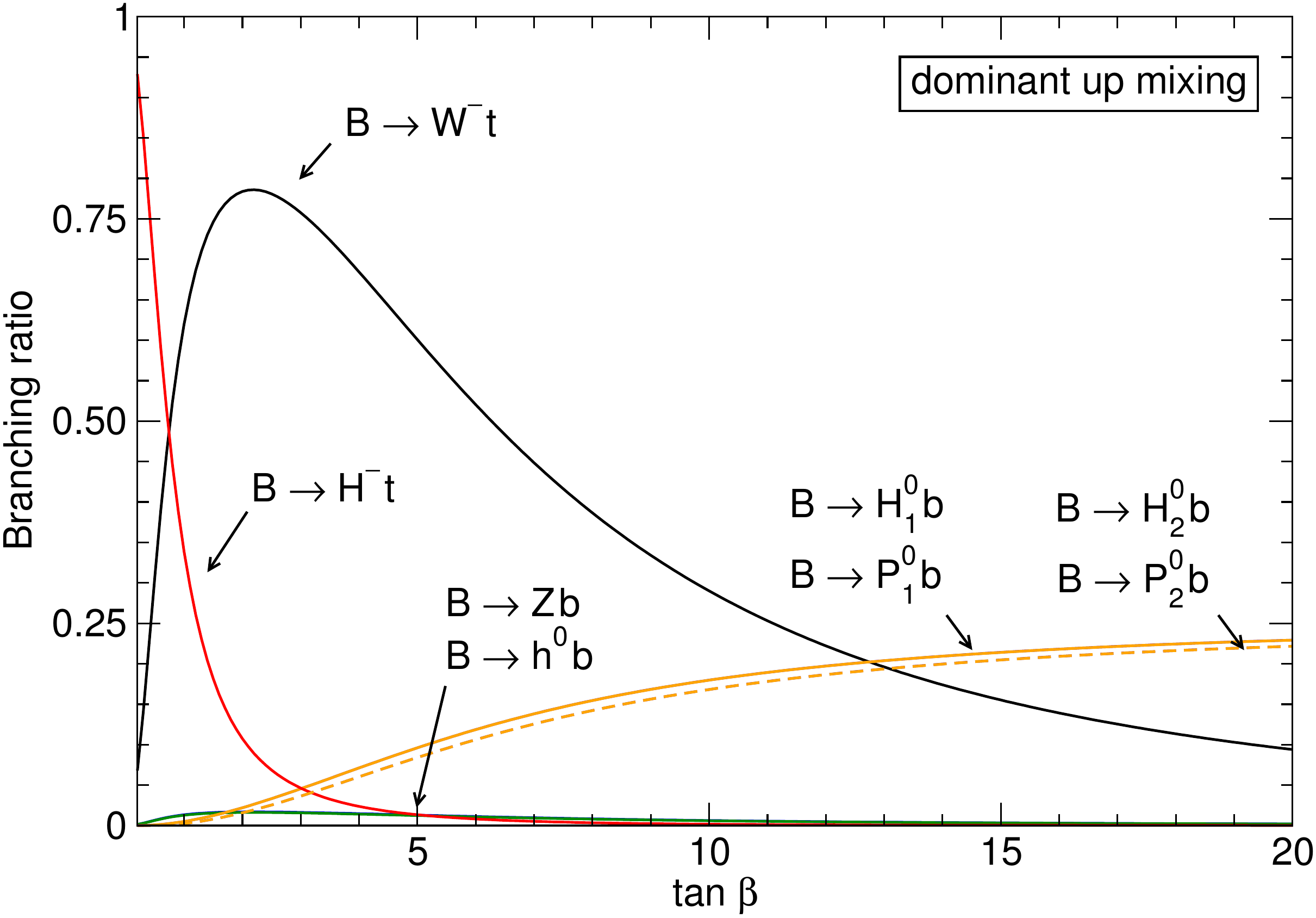} \\
\includegraphics[height=5.25cm,clip=]{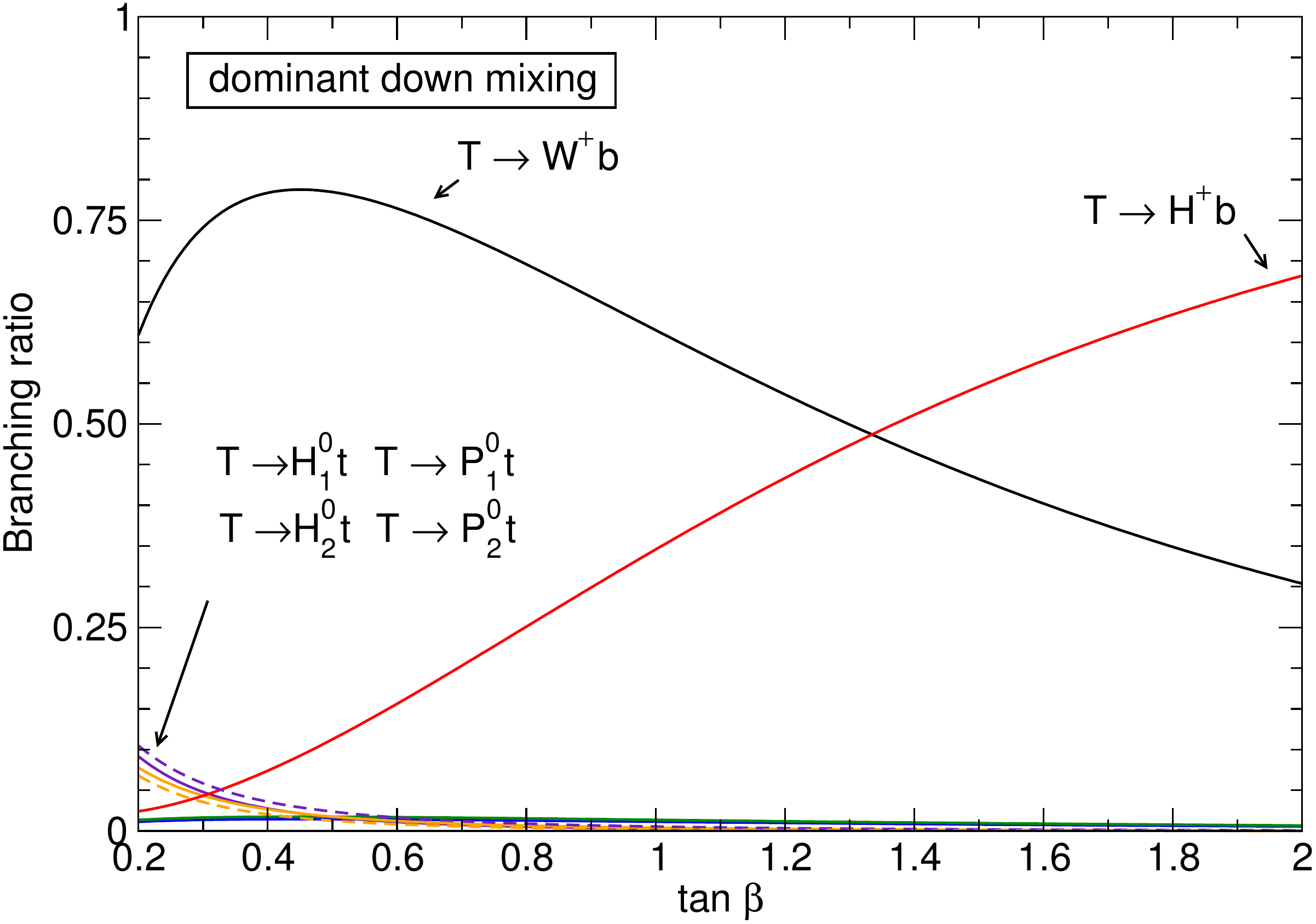} &
\includegraphics[height=5.25cm,clip=]{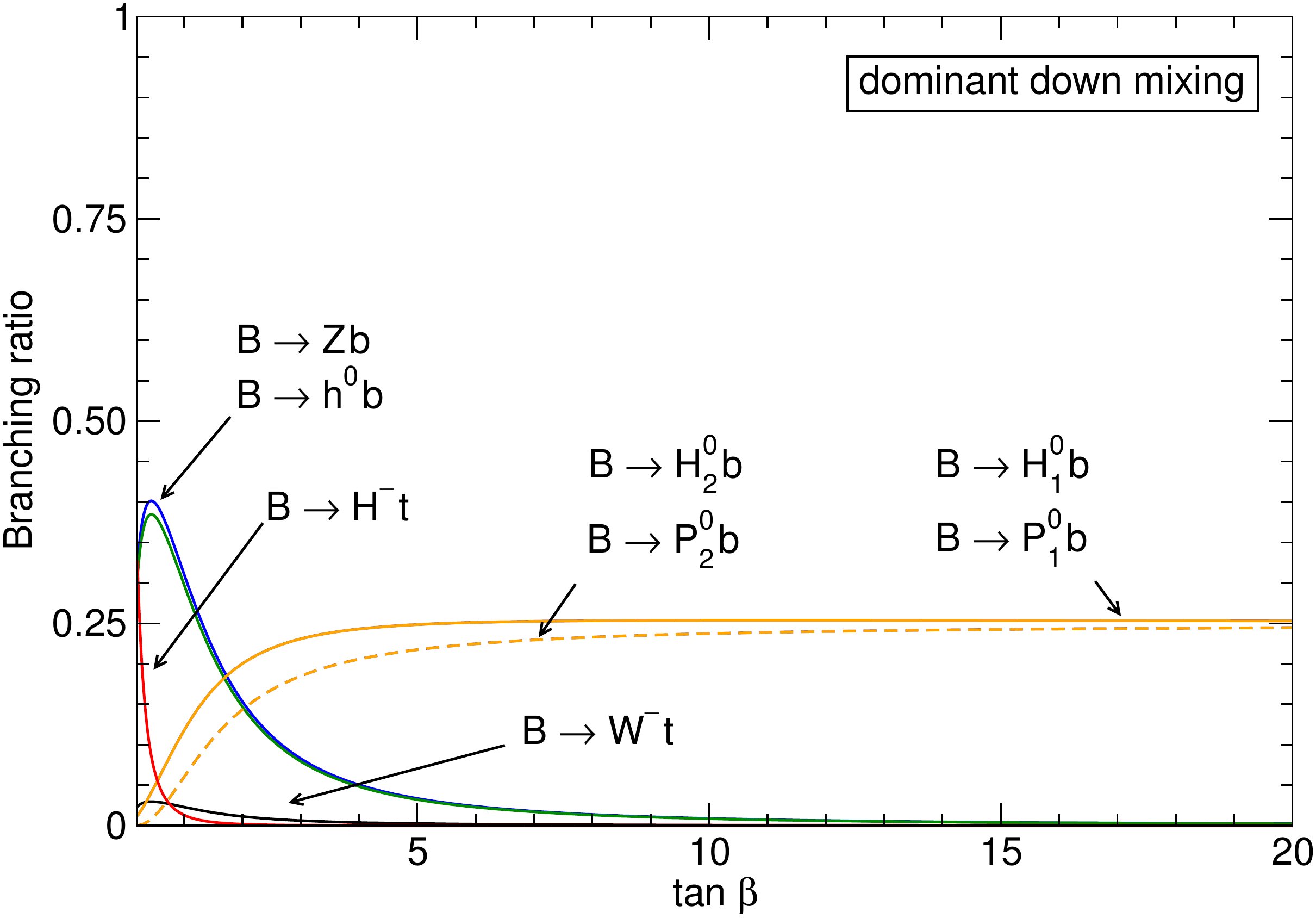}
\end{tabular}
\caption{Dependence of the $T$ and $B$ branching ratios on $\tan \beta$, for the equal mixing and dominant up/down mixing scenario of tables~\ref{tab:coup-eq}, \ref{tab:coup-up} and \ref{tab:coup-down}.}
\label{fig:BR3}
\end{center}
\end{figure}

For $T$ quark decays, we have found in section~\ref{sec:4.1}  that  the $H_1^0$ and $P_1^0$ modes are only relevant when $\tan \beta \lesssim 1$. In such case, the $1/\kappa_R$ terms in eqs. (\ref{ec:repl1}) are subdominant and the widths for $T \to H_1^0 t$ and $T \to P_1^0 t$ for the two scalar doublet model are shared with the additional modes, with weights proportional to the sine or cosine squared of the scalar mixing angles,
\begin{align}
&\Gamma(T \to H_1^0 t) \simeq  \left. \Gamma(T \to H_1^0 t)\right|_\text{2DM} \times \cos^2 \theta'  \,, \notag \\
&\Gamma(T \to H_2^0 t) \simeq  \left. \Gamma(T \to H_1^0 t)\right|_\text{2DM} \times \sin^2 \theta'  \,, \notag \\ \
& \Gamma(T \to P_1^0 t) \simeq  \left. \Gamma(T \to P_1^0 t) \right|_\text{2DM} \times \cos^2 \theta \,, \notag \\
& \Gamma(T \to P_2^0 t) \simeq  \left. \Gamma(T \to P_1^0 t) \right|_\text{2DM} \times \sin^2 \theta \,, 
\end{align}
up to small corrections from the $1/\kappa_R$ term and the possibly different scalar masses. We give some examples in figure~\ref{fig:BR3} (left), for the equal mixing and dominant up/down mixing scenarios, taking $\theta = \theta' = \pi/4$. For clarity, we zoom on the low $\tan \beta$ region. For the rest of mixing scenarios the results can easily be obtained from figures~\ref{fig:BR1} and \ref{fig:BR2} and the above equations.
For $B$ quark decays, we have also seen in section~\ref{sec:4.1} that decays to $H_1^0 b$ and $P_1^0 b$ are relevant only for $\tan \beta \gtrsim 1$. In this case, the $1/\kappa_R$ terms in eqs. (\ref{ec:repl2}) are subleading and we have
\begin{align}
&  \Gamma(B \to H_1^0 b) \simeq \left. \Gamma(B \to H_1^0 b) \right|_\text{2DM} \times \cos^2 \theta' \,, \notag \\ \displaybreak
&  \Gamma(B \to H_2^0 b) \simeq \left. \Gamma(B \to H_1^0 b) \right|_\text{2DM} \times \sin^2 \theta' \,, \notag \\
&  \Gamma(B \to P_1^0 b) \simeq \left. \Gamma(B \to P_1^0 b) \right|_\text{2DM}  \times \cos^2 \theta \,, \notag \\
&  \Gamma(B \to P_2^0 b) \simeq \left. \Gamma(B \to P_1^0 b) \right|_\text{2DM}  \times \sin^2 \theta \,.
\end{align}
Some examples, for the equal mixing and dominant up/down mixing scenarios, are shown in the right panel of figure~\ref{fig:BR3}. Notice that the branching ratios for $H_1^0$ and $P_1^0$ decays, and also for $H_2^0$ and $P_2^0$, almost coincide, in contrast with $T$ decays, because the interference terms proportional to $m_b$ are negligible. We remark that the branching ratios for the rest of modes are almost the same as in the two doublet model, precisely due to eqs. (\ref{ec:repl1}) and (\ref{ec:repl2}).

The decay widths of $H_k^0$ and $P_k^0$ into $t \bar t$ and $b \bar b$ can be obtained from the ones given in appendix~\ref{sec:a} with the replacements (\ref{ec:repl3}) and (\ref{ec:repl4}). They thus depend not only on $\tan \beta$ but on the scalar mixing angles $\theta$, $\theta'$ and the quark mixing. In addition, the scalars $H_k^0$ can decay into $h^0 h^0$, with partial widths that depend on independent parameters~\cite{Fidalgo:2011ky}, and the branching ratio for a mass eigenstate that is mostly a $\tilde \nu_R$ can be of order unity. Also, cascade decays are possible (note that here we have considered the same mass for all scalars, for simplicity, but this is not the general case), giving a variety of final states, whose detailed analysis is beyond the scope of this work.


\section{Connection to standard searches}
\label{sec:5}

\begin{figure}[htb]
\begin{center}
\begin{tabular}{cc}
\includegraphics[height=5.25cm,clip=]{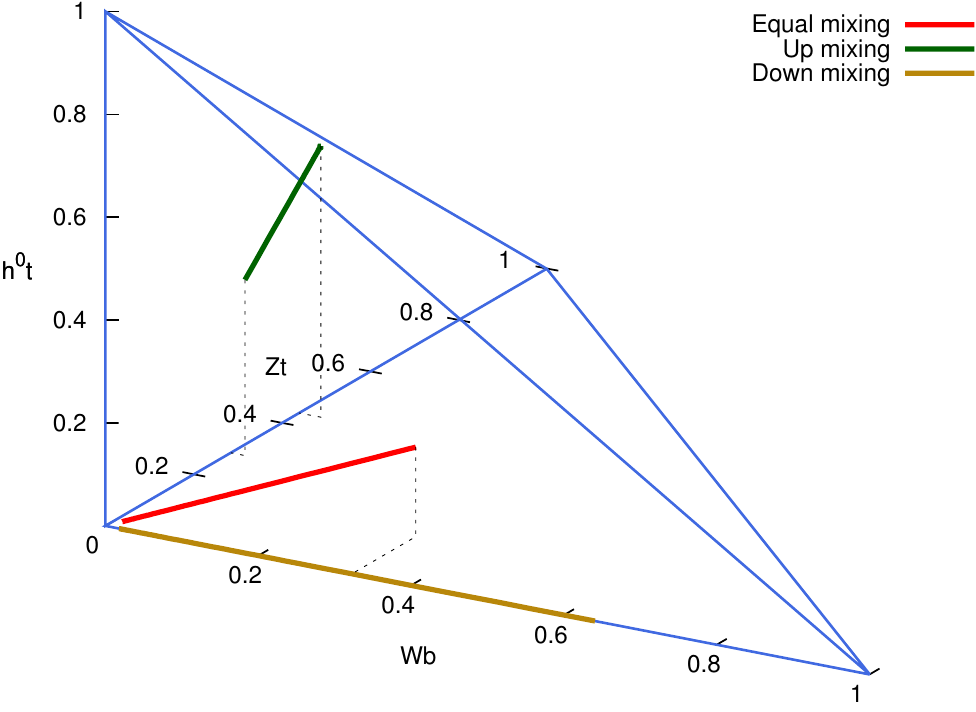} &
\includegraphics[height=5.25cm,clip=]{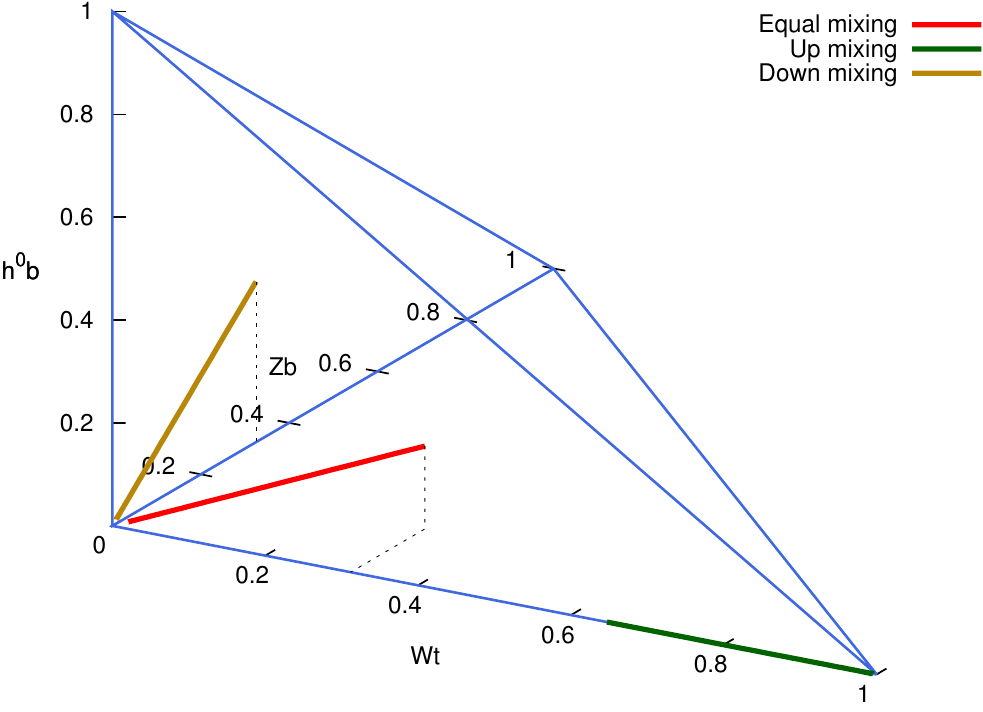}
\end{tabular}
\caption{Three-dimensional representation of the branching ratios to $W$, $Z$ and $h^0$ final states, for the $T$ quark (left) and the $B$ quark (right). }
\label{fig:3D}
\end{center}
\end{figure}

In the minimal SM extensions with a vector-like singlet, doublet or triplet and one scalar doublet the branching ratios of $T$ ($B$) decays to $Wb$ ($Wt$), $Zt$ ($Zb$) and $h^0 t$ ($h^0b$) add up to unity. We will refer to the branching ratios for these final states as $\text{Br}(W)$, $\text{Br}(Z)$ and $\text{Br}(h^0)$ when considering indistinctly $T$ and $B$ quarks. A given set of branching ratios, determined by the heavy quark masses and mixing parameters, can be represented in a triangle where two of the axes are, for example, $\text{Br}(Z)$ and $\text{Br}(h^0)$, and the third one is determined by the constraint that the sum equals one (see for example ref.~\cite{Aguilar-Saavedra:2013qpa}). This representation is also very convenient to give the result of experimental searches~\cite{Aad:2015mba,Aad:2015gdg,Khachatryan:2015gza,Aad:2015kqa}. In models with more than one scalar doublet this is no longer the case, and instead we have an inequality
\begin{equation}
\text{Br}(W) + \text{Br}(Z) + \text{Br}(h^0) \leq 1 \,.
\label{ec:ineq}
\end{equation}
A set of branching ratios to $W$, $Z$ and $h^0$ final states can then be represented by a point in three-dimensional space, within the pyramid obtained by the intersection of the coordinate planes and the plane $\text{Br}(W) + \text{Br}(Z) + \text{Br}(h^0) = 1$, as in figure~\ref{fig:3D}. Notice that the apex of the pyramid is the origin, and the pyramid is resting on a lateral face. Points in the equilateral triangle that is the base of the pyramid saturate the inequality (\ref{ec:ineq}), i.e. no decays into the new modes involving $H_k^0$, $P_k^0$ or $H^\pm$. As one approaches the origin, these new modes dominate and at the origin $\text{Br}(W) + \text{Br}(Z) + \text{Br}(h^0) = 0$. 
This graphical representation does not capture the different weights of the new modes ($H_k^0$, $P_k^0$ and $H^\pm$) but to have a unique correspondence we would need a polyhedron in five- or seven-dimensional space, which is difficult to draw on a two-dimensional plot. In any case, this representation is useful as the current searches precisely target the $W$, $Z$ and $h^0$ decay modes and  in principle have less sensitivity to the new ones.

We also plot in figure~\ref{fig:3D} the branching ratios of the $W$, $Z$ and $h^0$ decay modes for the heavy $T$ quark (left) and $B$ quark (right), with $\tan \beta$ ranging from 1 to 10, for the scenario with two scalar doublets and three quark mixing benchmarks. (In the model with an additional scalar singlet the results are very close due to eqs. (\ref{ec:repl1}) and (\ref{ec:repl2}).) The results are in agreement with those in figures~\ref{fig:BR1} and \ref{fig:BR2}:
\begin{itemize}
\item[(i)] In the equal mixing scenario (red), the points corresponding to $\tan \beta = 1$ are located in the interior of the pyramid and approach the origin as $\tan \beta$ increases.
\item[(ii)] In the only up mixing scenario (green), for the $T$ quark the $\tan \beta = 1$ point is inside the $\text{Br}(Wb) = 0$ lateral face and approaches the side $\text{Br}(Zt) + \text{Br}(h^0t) = 1$ as $\tan \beta$ increases; for the $B$ quark the point with $\tan \beta = 1$ is in the side $\text{Br}(Zb) = \text{Br}(h^0b) = 0$ and approaches the vertex $\text{Br}(Wt) = 1$ with increasing $\tan \beta$. 
\item[(iii)] In the only down mixing scenario (brown), for the $T$ quark the points move within the side $\text{Br}(Zt) = \text{Br}(h^0t) = 0$ towards the origin with increasing $\tan \beta$. For the $B$ quark the $\tan \beta$ point is inside the $\text{Br}(Wt) = 0$ lateral face and moves to the origin as $\tan \beta$ increases.
\end{itemize}

\begin{figure}[htb]
 \begin{center}
 \includegraphics[height=6cm,clip=]{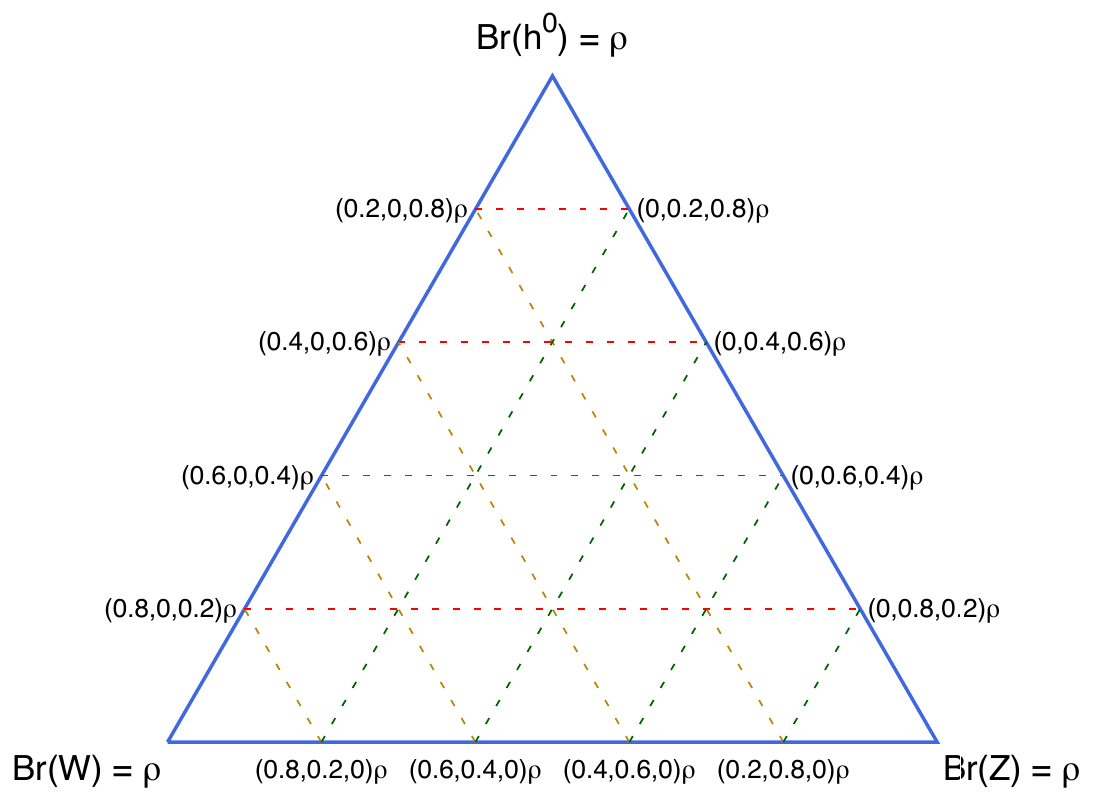}
 \caption{Two-dimensional slice of the pyramids in figure~\ref{fig:3D}, with $0 \leq \rho \leq 1$. Lines of constant $\text{Br}(W)$, $\text{Br}(Z)$ and $\text{Br}(h^0)$ are drawn in brown, green and red, respectively.}
 \label{fig:2D}
 \end{center}
 \end{figure}

Although the three-dimensional pyramids are convenient to represent the model predictions, they may not be useful to give the results of searches that cover the full volume. Instead, one can work with triangular slices parallel to the base,
\begin{equation}
\text{Br}(W) + \text{Br}(Z) + \text{Br}(h^0) = \rho \,,
\end{equation}
with $0 \leq \rho \leq 1$. One of such slices is represented in figure~\ref{fig:2D}, and is analogous to the triangles sometimes used by the CMS Collaboration to give the results of the heavy quark searches~\cite{Chatrchyan:2013uxa,Khachatryan:2015axa,Khachatryan:2015oba}. The three vertices of the triangle correspond to $\text{Br}(W)$, $\text{Br}(Z)$ or $\text{Br}(h^0)$ equal to $\rho$. The lines of constant $\text{Br}(h^0)$ are horizontal and the distance to the base is proportional to $\text{Br}(h^0)/\rho$. The same can be said about the lines of constant $\text{Br}(W)$ or $\text{Br}(Z)$: they are parallel to the opposite side of the triangle, and the distance to that side is proportional to $\text{Br}(W)/\rho$ or $\text{Br}(Z)/\rho$, respectively. For illustration, in figure~\ref{fig:2D} we mark several points with the values of $(\text{Br}(W),\text{Br}(Z),\text{Br}(h^0))$. The details concerning the correspondence of triangle points with branching ratios are given in appendix~\ref{sec:b}.

\section{Recasting searches: an example}
\label{sec:6}

Limits on heavy quark masses for arbitrary branching ratios into $W$, $Z$ and $h^0$ are obtained by the ATLAS and CMS Collaborations using the following procedure. For a given heavy quark mass, samples are generated corresponding to each pair $i,j$ of decay modes, with $i,j=W,Z,h^0$. The signal efficiency after event selection of each combination, which we denote as $\epsilon_{ij}$,  is calculated from simulation. Then, the efficiency for arbitrary branching ratios of the heavy quark can be written as
\begin{equation}
\epsilon = \sum_{i,j=W,Z,h^0} \epsilon_{ij} \; \text{Br}(i) \, \text{Br}(j) \,.
\end{equation}
Given this efficiency, and the measured limit on cross section times efficiency, the limits on cross section can be obtained, which can be translated into limits on the mass of the heavy quark, using the cross section predictions from the theory.

In the presence of new decay modes the procedure is the same, but extending the sum on $i,j$ over additional channels $X$. The results can be represented in the triangles introduced in the previous section, by making some hypothesis on these additional channels. The first possibility is to assume that the new decays are just invisible to the search\footnote{This not the same as assuming that the heavy quark decays invisibly.},  which corresponds to setting $\epsilon_{iX} = 0$, $\epsilon_{XX} = 0$. That gives a conservative limit on cross sections.
The second possibility is to make some hypothesis for the new decays and marginalise over these new degrees of freedom when interpreting the limits on the branching ratios of the standard modes. We will show an example of the latter.

 \begin{figure}[p]
 \begin{center}
 \begin{tabular}{cc}
 \includegraphics[height=5.5cm,clip=]{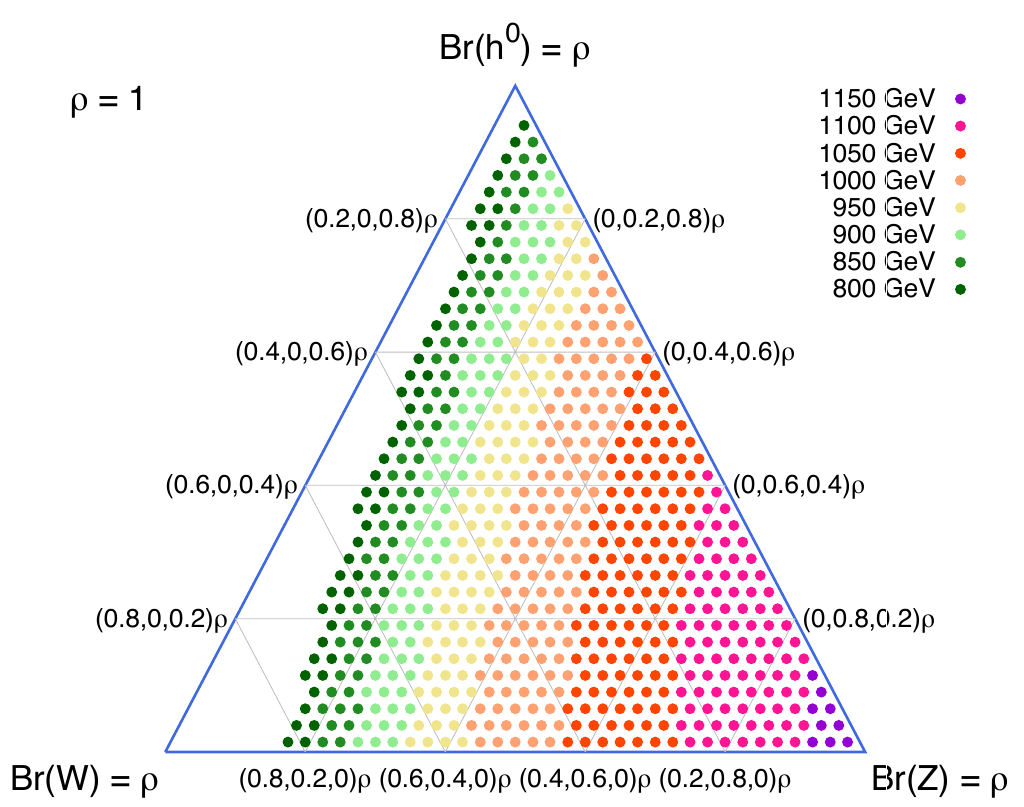} &
 \includegraphics[height=5.5cm,clip=]{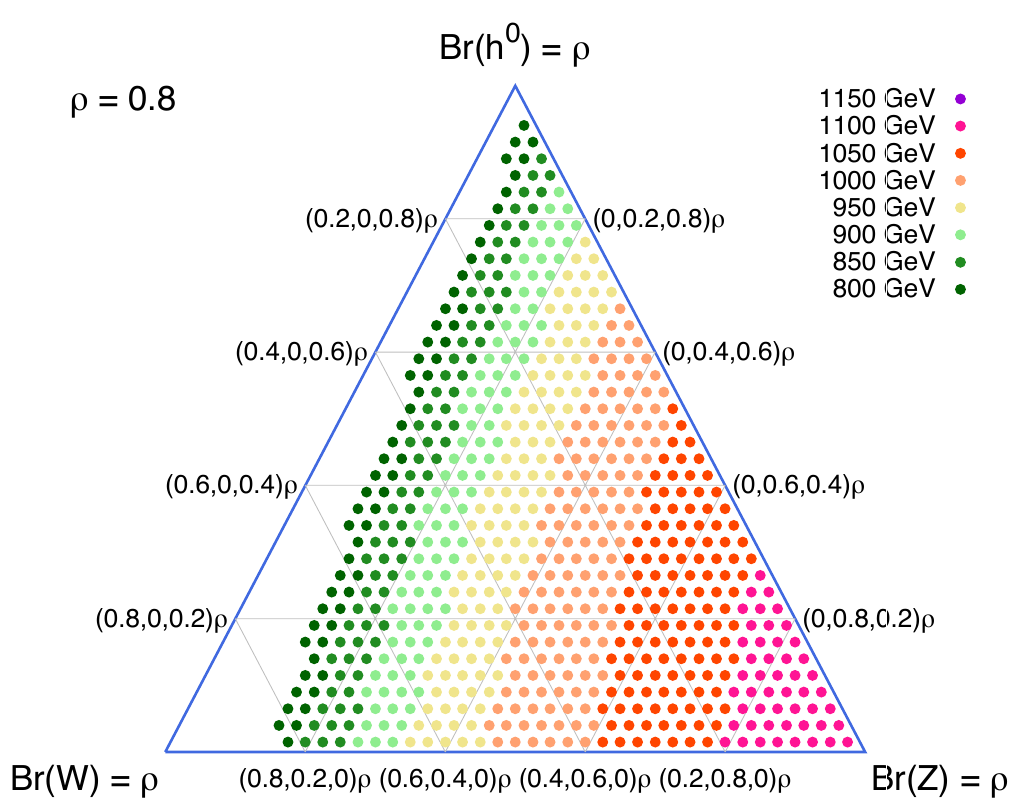} \\[2mm]
 \includegraphics[height=5.5cm,clip=]{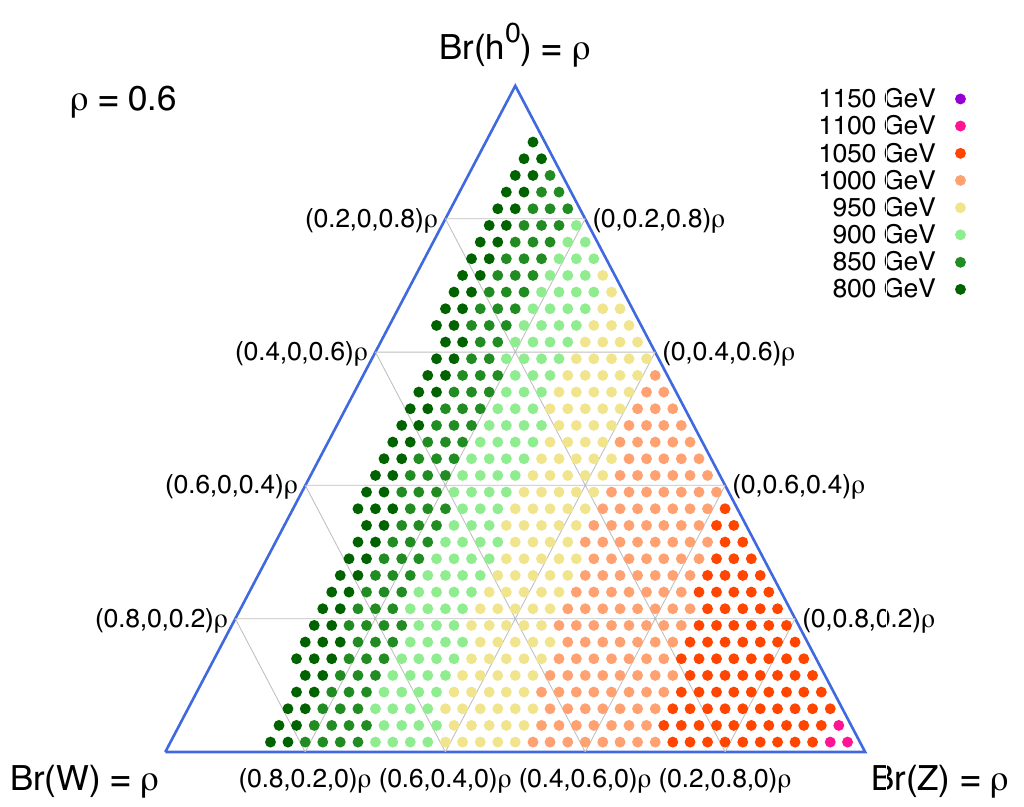} &
 \includegraphics[height=5.5cm,clip=]{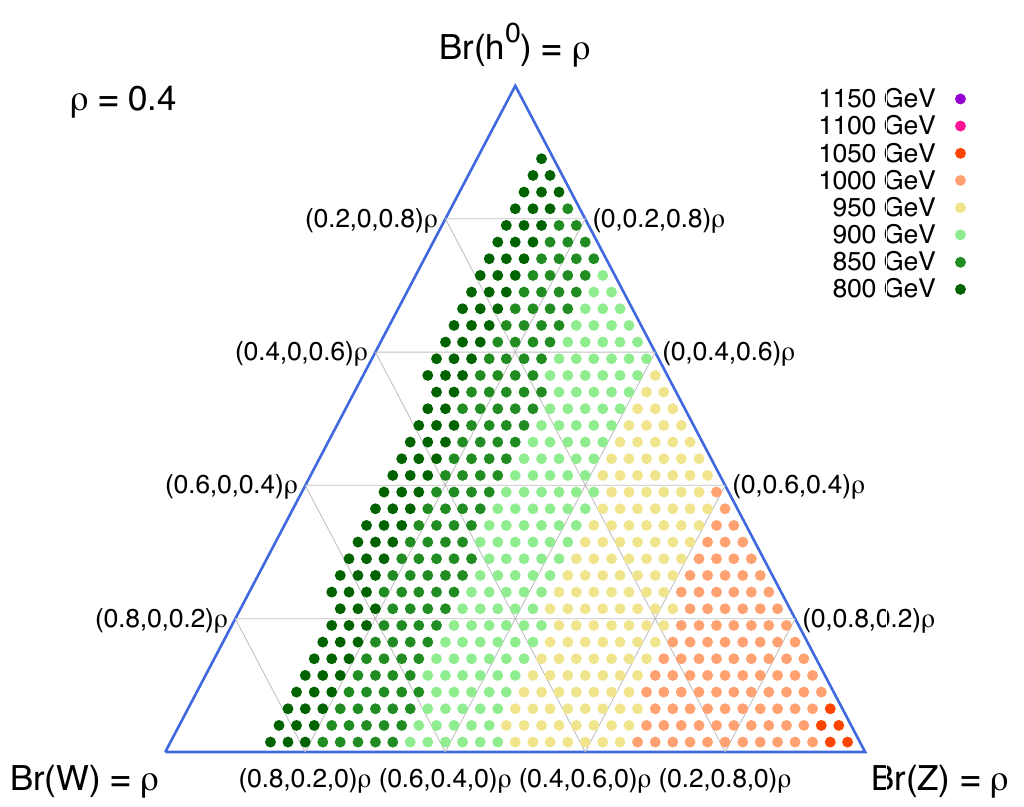} \\[2mm]
 \multicolumn{2}{c}{\includegraphics[height=5.5cm,clip=]{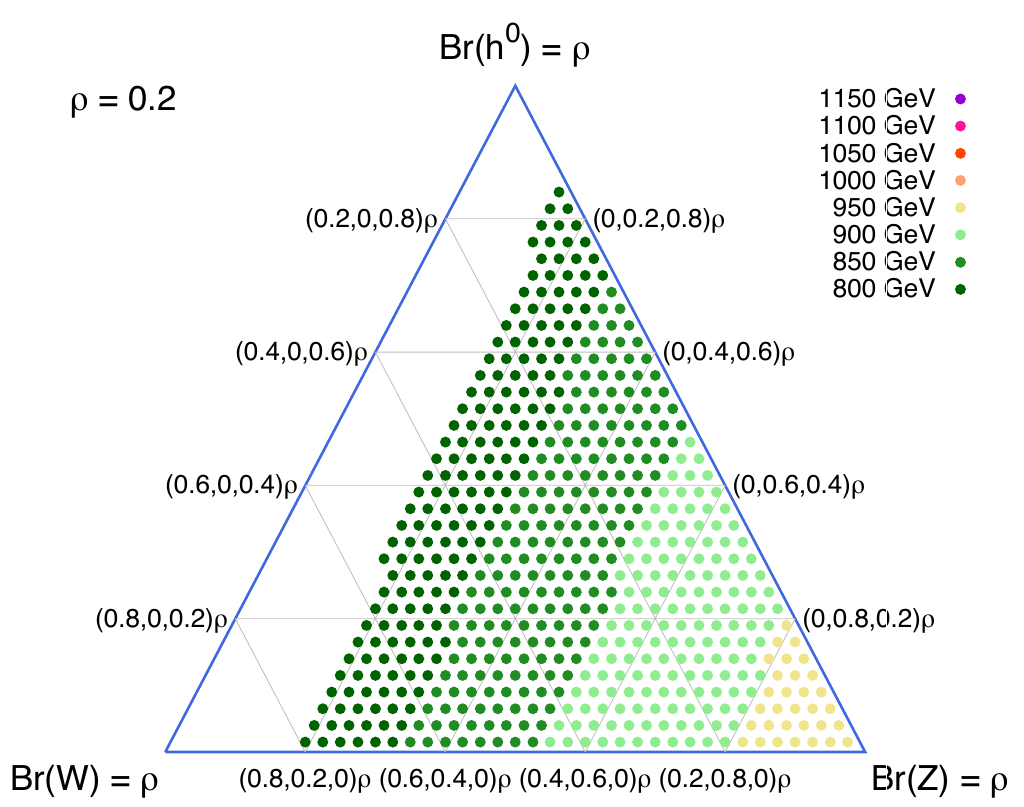}}
 \end{tabular}
 \caption{Lower limits on the $T$ quark mass for several values of $\rho$ (see the text), from a recast of the limits of the heavy quark search in ref.~\cite{confnote}.}
 \label{fig:limits}
 \end{center}
 \end{figure}

We consider the search for $T \bar T$ production in ref.~\cite{confnote} by the ATLAS Collaboration. This analysis focuses on final states with a single charged lepton, large missing energy (from the invisible decay of the $Z$ boson in $T \to Zt$), and at least four jets, reconstructed with the anti-$k_T$ algorithm~\cite{Cacciari:2008gp}, two of which are large-radius jets with $R = 1.0$, corresponding to boosted $W$ bosons. Small-radius jets with $R = 0.4$ are also used, requiring one of them to be $b$-tagged.
For a $T$ mass of 1 TeV, the relative fraction of events after event selection for the different decay modes of the $T \bar T$ pair assuming $\text{Br}(W) = \text{Br}(Z) = \text{Br}(h^0) = 1/3$ is given, as well as the global efficiency of $1\%$ for the benchmark point $\text{Br}(Z) = 0.8$, $\text{Br}(W) = \text{Br}(h^0) = 0.1$. These data allow us to extract the efficiencies for the different channels, relative to all decay modes of the top quarks and $W$, $Z$ bosons,
\begin{equation}
\epsilon = \left( \!\! \begin{array}{ccc}
6.9 \times 10^{-4} & 5.2 \times 10^{-3} &  1.6 \times 10^{-3} \\
5.2 \times 10^{-3} & 1.2 \times 10^{-2} &  9.0 \times 10^{-3} \\
1.6 \times 10^{-3} & 9.0 \times 10^{-3} & 1.8 \times 10^{-3}
\end{array} \!\! \right) \,,
\end{equation} 
where we have ordered the channels as $W$, $Z$, $h^0$.
We note in passing that this matrix has rank three, indicating that the efficiencies do not factorise, that is, they cannot be written as $\epsilon_{ij} = \hat \epsilon_i \hat \epsilon_j$. The search in ref.~\cite{confnote} is performed in a single event category, for which the expected number of background events is $6.1$, and the observed number of events is 7. For the luminosity of 36.1 fb$^{-1}$ used in that measurement, we can obtain a 95\% upper limit on the signal cross section using Feldman-Cousins confidence intervals~\cite{Feldman:1997qc}. The limits obtained are similar to those obtained by the ATLAS Collaboration with a profile likelihood using the $\text{CL}_s$ method~\cite{Read:2002hq}.

We recast the search by assuming that the efficiencies of decay modes involving the new channels are similar to the corresponding channels involving $T \to h^0 t$, that is,
$\epsilon_{iX} = \epsilon_{ih^0}$, $\epsilon_{XX} = \epsilon_{h^0 h^0}$. This is a well justified assumption for $\tan \beta \gtrsim 1$, because $T \to h^0 t \to b \bar b t$ and the new competing mode $T \to H^+ b \to t \bar b b$ lead to the same final state of an energetic top quark and two $b$ quarks. (In the sub-dominant decays $T \to H_k^0/P_k^0 t$ we would also have this final state as well as $t \bar t t$ and $h^0 h^0 t$.) We can then write the efficiency for arbitrary branching ratios, including the new modes, as
\begin{equation}
\epsilon = \sum_{i,j=W,Z,h^0,X} \epsilon_{ij} \; \text{Br}(i) \, \text{Br}(j) \,.
\end{equation}
With the so-calculated efficiency we obtain upper limits on the cross section, which we convert into lower limits on the new quark mass $m_T$, using the $T \bar T$ cross section at next-to-next-to-leading order~\cite{Czakon:2011xx}. We present our results in figure~\ref{fig:limits}, for several values of $\rho$ ranging from $1$ to $0.2$. For the standard case $\rho=1$ our limits are in very good agreement with the ones shown in ref.~\cite{confnote}. We do not present limits for  heavy quark masses lower than  $m_T = 800$ GeV because the efficiency changes with the mass, and for lower masses the approximation of taking the efficiency for $m_T = 1$ TeV may not be adequate.

\section{Discussion}
\label{sec:7}

In this paper we have considered a SM extension with an additional vector-like quark doublet and an extended scalar sector. The generalisation of the minimal vector-like models with one scalar doublet~\cite{delAguila:2000rc} is motivated by supersymmetric SM extensions, which at least have two Higgs doublets $H_u$ and $H_d$ as in the case of the MSSM. Supersymmetric models with extra scalar singlets are also possible. This is the case of the NMSSM where one extra singlet superfield is introduced to solve the $\mu$ problem.
In the case of the $\mu\nu$SSM~\cite{LopezFogliani:2005yw}, three families of right-handed neutrino superfields are used to solve this problem as well as to generate correct neutrino masses and mixing.
As a consequence, Higgses and right sneutrinos can have a sizeable mixing.

Compared to standard signatures~\cite{AguilarSaavedra:2009es}, in the models studied in this paper a wealth of new signals can be produced, among which we can mention the decays $T \to t \bar t t$, $T \to h^0 h^0 t$, $B \to t \bar t b$, $B \to h^0 h^0 b$, mediated by the new neutral scalars, as well as their combination with the standard modes, when the heavy quarks are produced in pairs. The vector-like doublet has two independent mixing angles in the up and down sectors, $s_R^u$ and $s_R^d$, respectively, which control the relative branching ratios of the charged and neutral decay modes. This freedom, together with the dependence of the new decay modes on the ratio of VEVs $\tan \beta$, gives rise to a large number of possibilities for the decay of the new quarks, which have been studied in detail in section~\ref{sec:4}. The decay modes of the $T$ and $B$ quarks in a vector-like doublet are the same that would be produced for other vector-like multiplets. In this sense, the model considered here is representative of the possible new signals for vector-like quark decays in supersymmetric models. 

A simple graphical representation of heavy quark branching ratios in three-dimensional pyramids has been given in section~\ref{sec:5}. For a generalised vector-like quark model, this allows us for example to see at a glance to which extent the collider signals of the heavy quarks $T$ and $B$ in the model  under consideration are close to the standard signals searched for by ATLAS and CMS --- and therefore covered by them. By using two-dimensional triangular slices, the same data can also be presented in two-dimensional plots. This latter representation is more convenient for the presentation of limits on quark masses from experimental searches, and is a generalisation of the equilateral triangles used by the CMS Collaboration.

Heavy quark searches carried out at the LHC cannot cover all the possibilities for the decays of the $T$ and $B$ quarks. But, under reasonable assumptions, the results of experimental searches in the standard decay modes can be interpreted in more general scenarios where the branching ratios to $W$, $Z$ and $h^0$ final states do not sum up to one, $\text{Br}(W) + \text{Br}(Z) + \text{Br}(h^0) < 1$, due to additional channels. One very rough and conservative assumption would be to take the new channels just as invisible for the searches. More refined interpretations can be made by evaluating the efficiencies of the event selection for the new channels, and marginalising the limits obtained over their branching ratios. We have provided one example in section~\ref{sec:6}, by recasting an existing search. Such generalised interpretations of experimental searches are feasible and should be pursued by the ATLAS and CMS Collaborations.

Finally, we point out that some of the new signatures like multi-top and multi-Higgs production are quite striking and worth exploring with dedicated searches. And, in any case, these new decay modes should be kept in mind when designing the event selection of the searches, to try to make them as inclusive as possible, and sensitive to these new signatures of heavy quarks.

\section*{Acknowledgements}
The work of J.A. Aguilar-Saavedra has been supported by MINECO Projects  FPA 2016-78220-C3-1-P and FPA 2013-47836-C3-2-P (including ERDF), and by Junta de Andaluc\'{\i}a Project FQM-101.
The work of D.E. L\'opez-Fogliani has been supported by the Argentinian CONICET.
The work of C. Mu\~noz has been supported in part 
by the Programme SEV-2012-0249 `Centro de Excelencia Severo Ochoa'.
D.E. L\'opez-Fogliani and C. Mu\~noz also acknowledge the support of the Spanish grant FPA2015-65929-P (MINECO/FEDER, UE), and MINECO's Consolider-Ingenio 2010 Programme under grant MultiDark CSD2009-00064.

\appendix
\section{Decay widths}
\label{sec:a}

The partial widths for the heavy quark decays into gauge or SM Higgs bosons are well known, and we collect them here for completeness, together with the partial widths for the new modes, in the case of no mixing with the scalar singlet. When the mixing is significant, the corresponding expressions can be obtained from the ones below with the replacements (\ref{ec:repl1}) and (\ref{ec:repl2}).

Defining $r_x = m_x / m_Q$, where $Q$ is the heavy quark and $x$ one of its decay products, and the function
\begin{equation}
\lambda(x,y,z) \equiv (x^4 + y^4 + z^4 - 2 x^2 y^2 
- 2 x^2 z^2 - 2 y^2 z^2) \,,
\end{equation}%
the partial widths for $T$ decays are
\begin{align}
\Gamma(T \to W^+ b) & = \frac{g^2}{64 \pi}  \frac{m_T}{M_W^2} \lambda(m_T,m_b,M_W)^{1/2} \left\{
(|V_{Tb}^L|^2+|V_{Tb}^R|^2)  \right. \nonumber \\ 
  & \left. \times \left[ 1+r_W^2-2 r_b^2 -2 r_W^4  + r_b^4 +r_W^2 r_b^2
  \right]  -12 r_W^2 r_b \RE V_{Tb}^L V_{Tb}^{R*} \right\}
\,, \nonumber \\ 
\Gamma(T \to Z t) & = \frac{g^2}{128 \pi c_W^2}  \frac{m_T}{M_Z^2} \lambda(m_T,m_t,M_Z)^{1/2}
\left\{ (|X_{tT}^L|^2 + |X_{tT}^R|^2) \right. \nonumber \\
  & \left. \times  \left[ 1 + r_Z^2 - 2  r_t^2 - 2  r_Z^4  + r_t^4
  + r_Z^2 r_t^2 \right]  -12 r_Z^2 r_t \RE X_{tT}^L X_{tT}^{R*}  \right\} \,, \nonumber \\ 
\Gamma(T \to h^0 t) & = \frac{g^2}{128 \pi}
 \frac{m_T}{M_W^2} \lambda(m_T,m_t,M_{h^0})^{1/2}
 \left\{ (|Y_{tT}^L|^2 + |Y_{tT}^R|^2) \left[1+r_t^2 - r_{h^0}^2 \right] \right. \notag \\
 & \left. + 4 r_t \RE Y_{tT}^L Y_{tT}^{R*} \right\} \,, \notag \\ 
\Gamma(T \to H_1^0 t) & = \frac{g^2 \cot^2 \beta}{128 \pi} 
 \frac{m_T}{M_W^2} \lambda(m_T,m_t,M_{H_1^0})^{1/2}
 \left\{ (|Y_{tT}^L|^2 + |Y_{tT}^R|^2) \left[1+r_t^2 - r_{H_1^0}^2 \right] \right. \notag \\
 & \left. + 4 r_t \RE Y_{tT}^L Y_{tT}^{R*} \right\} \,, \notag \\ 
\Gamma(T \to P_1^0 t) & = \frac{g^2 \cot^2 \beta}{128 \pi} 
 \frac{m_T}{M_W^2} \lambda(m_T,m_t,M_{P_1^0})^{1/2}
 \left\{ (|Y_{tT}^L|^2 + |Y_{tT}^R|^2) \left[1+r_t^2 - r_{P_1^0}^2 \right] \right. \notag \\  \displaybreak
 & \left. - 4 r_t \RE Y_{tT}^L Y_{tT}^{R*} \right\} \,, \notag \\ 
\Gamma(T \to H^+ b) & = \frac{g^2 }{64 \pi} 
 \frac{m_T}{M_W^2} \lambda(m_T,m_b,M_{H^\pm})^{1/2}
 \left\{ (|Z_{Tb}^L|^2 \cot^2 \beta + |Z_{Tb}^R|^2 \tan^2 \beta )\right. \notag \\
 & \left. \times  \left[1+r_b^2 - r_{H^\pm}^2 \right]  + 4 r_b \RE Z_{Tb}^L Z_{Tb}^{R*} \right\} \,.
\label{ec:GammaT}
\end{align}
For the $B$ quark they are analogous,
\begin{align}
\Gamma(B \to W^- t) & = \frac{g^2}{64 \pi}  \frac{m_B}{M_W^2} \lambda(m_B,m_t,M_W)^{1/2} \left\{
(|V_{tB}^L|^2+|V_{tB}^R|^2)  \right. \nonumber \\
  & \left. \times \left[ 1+r_W^2-2 r_t^2  -2 r_W^4  + r_t^4 +r_W^2 r_t^2
  \right]  -12 r_W^2 r_t \RE V_{tB}^L V_{tB}^{R*} \right\}
\,, \nonumber \\
\Gamma(B \to Z b) & = \frac{g^2}{128 \pi c_W^2}  \frac{m_B}{M_Z^2} \lambda(m_B,m_b,M_Z)^{1/2}
\left\{ (|X_{bB}^L|^2 + |X_{bB}^R|^2) \right. \nonumber \\
  & \left. \times  \left[ 1 + r_Z^2 - 2  r_b^2 - 2  r_Z^4  + r_b^4
  + r_Z^2 r_b^2 \right]  -12 r_Z^2 r_b \RE X_{bB}^L X_{bB}^{R*}  \right\} \,, \nonumber \\
\Gamma(B \to h^0 b) & = \frac{g^2}{128 \pi}
 \frac{m_B}{M_W^2} \lambda(m_B,m_b,M_{h^0})^{1/2}
 \left\{ (|Y_{bB}^L|^2 + |Y_{bB}^R|^2) \left[1+r_b^2 - r_{h^0}^2 \right] \right. \notag \\
 & \left. + 4 r_b \RE Y_{bB}^L Y_{bB}^{R*} \right\} \,, \notag \\
\Gamma(B \to H_1^0 b) & = \frac{g^2 \tan^2 \beta}{128 \pi} 
 \frac{m_B}{M_W^2} \lambda(m_B,m_b,M_{H_1^0})^{1/2}
 \left\{ (|Y_{bB}^L|^2 + |Y_{bB}^R|^2) \left[1+r_b^2 - r_{H_1^0}^2 \right] \right. \notag \\ 
 & \left. + 4 r_b \RE Y_{bB}^L Y_{bB}^{R*} \right\} \,, \notag \\ 
\Gamma(B \to P_1^0 b) & = \frac{g^2 \tan^2 \beta}{128 \pi} 
 \frac{m_B}{M_W^2} \lambda(m_B,m_b,M_{P_1^0})^{1/2}
 \left\{ (|Y_{bB}^L|^2 + |Y_{bB}^R|^2) \left[1+r_b^2 - r_{P_1^0}^2 \right] \right. \notag \\ 
 & \left. - 4 r_b \RE Y_{bB}^L Y_{bB}^{R*} \right\} \,, \notag \\
\Gamma(B \to H^- t) & = \frac{g^2 }{64 \pi} 
 \frac{m_B}{M_W^2} \lambda(m_B,m_t,M_{H^\pm})^{1/2}
 \left\{ (|Z_{tB}^L|^2 \cot^2 \beta + |Z_{tB}^R|^2 \tan^2 \beta )\right. \notag \\
 & \left. \times  \left[1+r_t^2 - r_{H^\pm}^2 \right]  + 4 r_t \RE Z_{tB}^L Z_{tB}^{R*} \right\} \,.
\label{ec:GammaB}
\end{align}
The partial widths for the decays of $S^0 = H_1^0, P_1^0$ into $t \bar t$ and $b \bar b$ are
\begin{align}
& \Gamma(S^0 \to t \bar t) = \frac{N_c \, g^2}{32\pi} \frac{m_t^2}{M_W^2} Y_{tt}^2 \cot^2 \beta M_{S^0} \left[ 1-\frac{4 m_t^2}{M_{S^0}^2} \right]^p \,, \notag \\
& \Gamma(S^0 \to b \bar b) = \frac{N_c \, g^2}{32\pi} \frac{m_b^2}{M_W^2} Y_{bb}^2 \tan^2 \beta M_{S^0} \left[ 1-\frac{4 m_b^2}{M_{S^0}^2} \right]^p \,,
\end{align}
with $N_c = 3$ the number of colours and $p = 3/2$ ($1/2$) for $H_1^0$ ($P_1^0$).

\section{Geometry of the branching ratio triangles}
\label{sec:b}

\begin{figure}[htb]
 \begin{center}
 \includegraphics[height=4cm,clip=]{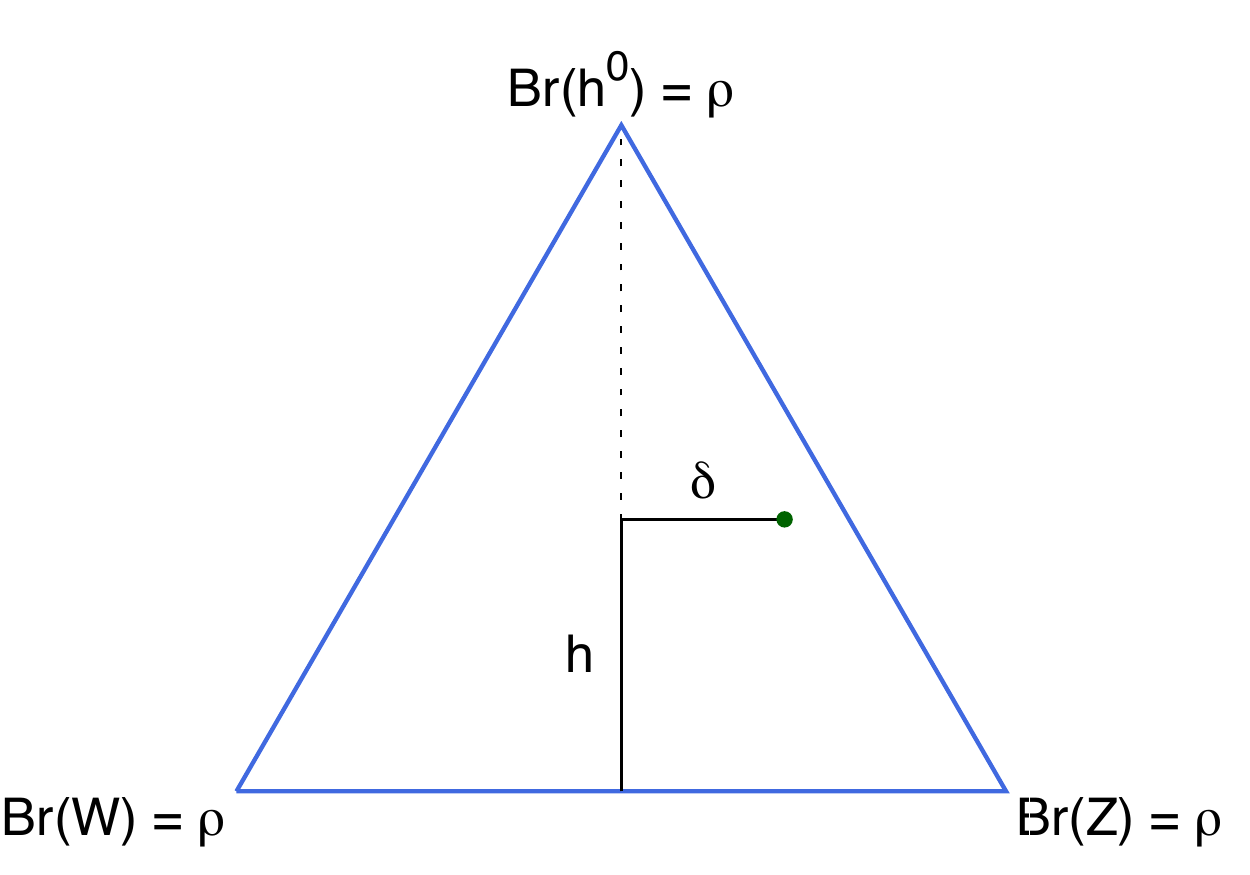}
 \caption{Coordinates used to parameterise branching ratios in a triangle.}
 \label{fig:coord}
 \end{center}
 \end{figure}

In the equilateral triangle of figure~\ref{fig:coord}, representing the slice of the pyramid with $\text{Br}(W) + \text{Br}(Z) + \text{Br}(h^0) = \rho$, one can introduce two coordinates $h$ and $\delta$, with $h \in [0,\sqrt{3/2} \rho]$ corresponding to the height over the base and $\delta \in [-(\rho/\sqrt 2 - h/\sqrt 3),\rho/\sqrt 2 - h/\sqrt 3]$ the horizontal displacement from the vertical median.
In terms of them, the branching ratios are
\begin{align}
& \text{Br}(W) = \frac{\rho}{2} - \frac{h}{\sqrt 6} - \frac{\delta}{\sqrt 2} \,, \quad
\text{Br}(Z) = \frac{\rho}{2} - \frac{h}{\sqrt 6} + \frac{\delta}{\sqrt 2} \,, \quad
\text{Br}(h^0) = \sqrt\frac{2}{3} h \,.
\label{ec:WZH-hdelta}
\end{align}
Conversely, $\delta = \text{Br}(Z) - \text{Br}(W)$, $h= \sqrt{3/2} \;  \text{Br}(h^0)$. The vertical median corresponds to $\text{Br}(W) = \text{Br}(Z)$, and the other two medians to $\text{Br}(W) = \text{Br}(h^0)$ and $\text{Br}(Z) = \text{Br}(h^0)$. The centroid has the three branching ratios equal to $\rho/3$. 
The lines of constant $\text{Br}(h^0)$  are horizontal,
\begin{equation}
h = \sqrt\frac{3}{2} \text{Br}(h^0) \,,
\end{equation}
as said before. The extreme points are at $\delta = \pm \left[ \rho-\text{Br}(h^0) \right]/\sqrt 2$.
The lines of constant $\text{Br}(W)$ are given by
\begin{equation}
h + \sqrt 3 \delta= \sqrt 6  \left[ \frac{\rho}{2} - \text{Br}(W) \right] \,,
\end{equation}
with extreme points at $h=0$, $\delta = \sqrt 2 \left[ \rho/2-\text{Br}(W) \right]$ and $h=\sqrt{3/2} \left[ \rho-\text{Br}(W) \right]$, $\delta = - \text{Br}(W) / \sqrt 2$. Lines of constant $\text{Br}(Z)$ are given by
\begin{equation}
h - \sqrt 3 \delta = \sqrt 6  \left( \frac{\rho}{2} - \text{Br}(Z) \right) \,,
\end{equation}
with extreme points at $h=0$, $\delta = - \sqrt 2 \left[ \rho/2-\text{Br}(Z) \right]$ and $h=\sqrt{3/2} \left[ \rho-\text{Br}(Z) \right]$, $\delta =\text{Br}(Z) / \sqrt 2$.


\begin{thebibliography}{99}

\bibitem{Perelstein:2003wd}
  M.~Perelstein, M.~E.~Peskin and A.~Pierce,
  Phys.\ Rev.\ {\bf D69} (2004) 075002
  [hep-ph/0310039].

\bibitem{Schmaltz:2005ky}
  M.~Schmaltz and D.~Tucker-Smith,
  Ann.\ Rev.\ Nucl.\ Part.\ Sci.\  {\bf 55} (2005) 229
  [hep-ph/0502182].

\bibitem{Contino:2006qr}
  R.~Contino, L.~Da Rold and A.~Pomarol,
  Phys.\ Rev.\ {\bf D75} (2007) 055014
  [hep-ph/0612048].

\bibitem{Matsedonskyi:2012ym}
  O.~Matsedonskyi, G.~Panico and A.~Wulzer,
  JHEP {\bf 01} (2013) 164
  [arXiv:1204.6333 [hep-ph]].

\bibitem{Contino:2006nn}
  R.~Contino, T.~Kramer, M.~Son and R.~Sundrum,
  JHEP {\bf 05} (2007) 074
  [hep-ph/0612180].

\bibitem{Kaplan:1991dc}
  D.~B.~Kaplan,
  Nucl.\ Phys.\ {\bf B365} (1991) 259.

\bibitem{Martin:1997ns}
For a review, see:  S.~P.~Martin,
  Adv.\ Ser.\ Direct.\ High Energy Phys.\  {\bf 21} (2010) 1
  [hep-ph/9709356].

\bibitem{Moroi:1991mg}
  T.~Moroi and Y.~Okada,
  Mod.\ Phys.\ Lett.\ A {\bf 7} (1992) 187.

\bibitem{Moroi:1992zk}
  T.~Moroi and Y.~Okada,
  Phys.\ Lett.\ B {\bf 295} (1992) 73.

\bibitem{Babu:2004xg}
  K.~S.~Babu, I.~Gogoladze and C.~Kolda,
  hep-ph/0410085.

\bibitem{Babu:2008ge}
  K.~S.~Babu, I.~Gogoladze, M.~U.~Rehman and Q.~Shafi,
  Phys.\ Rev.\ D {\bf 78} (2008) 055017
  [arXiv:0807.3055 [hep-ph]].

\bibitem{Martin:2009bg}
  S.~P.~Martin,
  Phys.\ Rev.\ D {\bf 81} (2010) 035004
  [arXiv:0910.2732 [hep-ph]].

\bibitem{Martin:2010dc}
  S.~P.~Martin,
  Phys.\ Rev.\ D {\bf 82} (2010) 055019
  [arXiv:1006.4186 [hep-ph]].
 
\bibitem{Graham:2009gy}
  P.~W.~Graham, A.~Ismail, S.~Rajendran and P.~Saraswat,
  Phys.\ Rev.\ D {\bf 81} (2010) 055016
  [arXiv:0910.3020 [hep-ph]].

\bibitem{Endo:2011mc}
  M.~Endo, K.~Hamaguchi, S.~Iwamoto and N.~Yokozaki,
  Phys.\ Rev.\ D {\bf 84} (2011) 075017
  doi:10.1103/PhysRevD.84.075017
  [arXiv:1108.3071 [hep-ph]].

\bibitem{Martin:2012dg}
  S.~P.~Martin and J.~D.~Wells,
  Phys.\ Rev.\ D {\bf 86} (2012) 035017
  [arXiv:1206.2956 [hep-ph]].

\bibitem{Faroughy:2014oka}
  C.~Faroughy and K.~Grizzard,
  Phys.\ Rev.\ D {\bf 90} (2014) no.3,  035024
  [arXiv:1405.4116 [hep-ph]].
  
\bibitem{Ellis:2014dza}
  S.~A.~R.~Ellis, R.~M.~Godbole, S.~Gopalakrishna and J.~D.~Wells,
  JHEP {\bf 1409} (2014) 130
  [arXiv:1404.4398 [hep-ph]].

\bibitem{Lalak:2015xea}
  Z.~Lalak, M.~Lewicki and J.~D.~Wells,
  Phys.\ Rev.\ D {\bf 91} (2015) no.9,  095022
  [arXiv:1502.05702 [hep-ph]].

\bibitem{Nickel:2015dna}
  K.~Nickel and F.~Staub,
  JHEP {\bf 1507} (2015) 139
  [arXiv:1505.06077 [hep-ph]].

\bibitem{Collins:2015wua}
  J.~H.~Collins and W.~H.~Ng,
  JHEP {\bf 1601} (2016) 159
  [arXiv:1510.08083 [hep-ph]].

\bibitem{Lopez-Fogliani:2017qzj}
  D.~E.~Lopez-Fogliani and C.~Mu\~noz,
  arXiv:1701.02652 [hep-ph], to appear in Phys. Lett. B.


\bibitem{LopezFogliani:2005yw}
  D.~E.~Lopez-Fogliani and C.~Mu\~noz,
  Phys.\ Rev.\ Lett.\  {\bf 97} (2006) 041801
  [hep-ph/0508297].



\bibitem{AguilarSaavedra:2009es}
  J.~A.~Aguilar-Saavedra,
  JHEP {\bf 11} (2009) 030
  [arXiv:0907.3155 [hep-ph]].


\bibitem{Aguilar-Saavedra:2013qpa}
  J.~A.~Aguilar-Saavedra, R.~Benbrik, S.~Heinemeyer and M.~P\'erez-Victoria,
  Phys.\ Rev.\ {\bf D88} (2013) no.9, 094010
  [arXiv:1306.0572 [hep-ph]].
  
\bibitem{Ellwanger:2009dp}
  U.~Ellwanger, C.~Hugonie and A.~M.~Teixeira,
  Phys.\ Rept.\  {\bf 496} (2010) 1
  [arXiv:0910.1785 [hep-ph]].

\bibitem{Kim:1983dt}
  J.~E.~Kim and H.~P.~Nilles,
  Phys.\ Lett.\  {\bf 138B} (1984) 150.

\bibitem{Munoz:2016vaa}
For a recent review, see:
  C.~Mu\~noz,
Proceedings of the 11th International
  Workshop on the Dark Side of the Universe (DSU2015), Kyoto, Japan, December
  14-18, 2015,  
  PoS DSU {\bf 2015} (2016) 034
  [arXiv:1608.07912 [hep-ph]].

\bibitem{Escudero:2008jg}
  N.~Escudero, D.~E.~Lopez-Fogliani, C.~Mu\~noz and R.~Ruiz de Austri,
  JHEP {\bf 0812} (2008) 099
  [arXiv:0810.1507 [hep-ph]].

\bibitem{Ghosh:2008yh}
  P.~Ghosh and S.~Roy,
  JHEP {\bf 0904} (2009) 069
  [arXiv:0812.0084 [hep-ph]].

\bibitem{Bartl:2009an}
  A.~Bartl, M.~Hirsch, A.~Vicente, S.~Liebler and W.~Porod,
  JHEP {\bf 0905} (2009) 120
  [arXiv:0903.3596 [hep-ph]].

\bibitem{Fidalgo:2009dm}
  J.~Fidalgo, D.~E.~Lopez-Fogliani, C.~Mu\~noz and R.~Ruiz de Austri,
  JHEP {\bf 0908} (2009) 105
  [arXiv:0904.3112 [hep-ph]].

\bibitem{Ghosh:2010zi}
  P.~Ghosh, P.~Dey, B.~Mukhopadhyaya and S.~Roy,
  JHEP {\bf 1005} (2010) 087
  [arXiv:1002.2705 [hep-ph]].

\bibitem{LopezFogliani:2010bf}
For a review, see:
  D.~E.~Lopez-Fogliani,
CTP International Conference on Neutrino Physics in the LHC Era, Luxor, Egypt,
  November 15-19, 2009 [arXiv:1004.0884[hep-ph]].





\bibitem{Chatrchyan:2013uxa}
  S.~Chatrchyan {\it et al.} [CMS Collaboration],
  Phys.\ Lett.\ {\bf B729} (2014) 149
  [arXiv:1311.7667 [hep-ex]].

\bibitem{Aad:2014efa}
  G.~Aad {\it et al.} [ATLAS Collaboration],
  JHEP {\bf 11} (2014) 104
  [arXiv:1409.5500 [hep-ex]].

\bibitem{Khachatryan:2015axa}
  V.~Khachatryan {\it et al.} [CMS Collaboration],
  JHEP {\bf 06} (2015) 080
  [arXiv:1503.01952 [hep-ex]].

\bibitem{Aad:2015mba}
  G.~Aad {\it et al.} [ATLAS Collaboration],
  Phys.\ Rev.\ {\bf D91} (2015) no.11,  112011
  [arXiv:1503.05425 [hep-ex]].

\bibitem{Aad:2015gdg}
  G.~Aad {\it et al.} [ATLAS Collaboration],
  JHEP {\bf 10} (2015) 150
  [arXiv:1504.04605 [hep-ex]].

\bibitem{Aad:2015kqa}
  G.~Aad {\it et al.} [ATLAS Collaboration],
  JHEP {\bf 08} (2015) 105
  [arXiv:1505.04306 [hep-ex]].

\bibitem{Khachatryan:2015gza}
  V.~Khachatryan {\it et al.} [CMS Collaboration],
  Phys.\ Rev.\ D {\bf 93} (2016) no.11,  112009
  [arXiv:1507.07129 [hep-ex]].


\bibitem{Khachatryan:2015oba}
  V.~Khachatryan {\it et al.} [CMS Collaboration],
  Phys.\ Rev.\  {\bf D93} (2016) no.1,  012003
  [arXiv:1509.04177 [hep-ex]].






\bibitem{Khachatryan:2015mta}
  V.~Khachatryan {\it et al.} [CMS Collaboration],
  JHEP {\bf 01} (2016) 166
  [arXiv:1509.08141 [hep-ex]].

\bibitem{Aad:2015voa}
  G.~Aad {\it et al.} [ATLAS Collaboration],
  JHEP {\bf 02} (2016) 110
  [arXiv:1510.02664 [hep-ex]].

\bibitem{Aad:2016qpo}
  G.~Aad {\it et al.} [ATLAS Collaboration],
  Eur.\ Phys.\ J.\  {\bf C76} (2016) no.8,  442
  [arXiv:1602.05606 [hep-ex]].

\bibitem{Khachatryan:2016vph}
  V.~Khachatryan {\it et al.} [CMS Collaboration],
  [arXiv:1612.00999 [hep-ex]].

\bibitem{Sirunyan:2017ezy}
  A.~M.~Sirunyan {\it et al.} [CMS Collaboration],
  arXiv:1701.07409 [hep-ex].

\bibitem{Sirunyan:2017tfc}
  A.~M.~Sirunyan {\it et al.} [CMS Collaboration],
  arXiv:1701.08328 [hep-ex].





\bibitem{Serra:2015xfa}
  J.~Serra,
  JHEP {\bf 1509} (2015) 176
  [arXiv:1506.05110 [hep-ph]].
  
\bibitem{Anandakrishnan:2015yfa}
  A.~Anandakrishnan, J.~H.~Collins, M.~Farina, E.~Kuflik and M.~Perelstein,
  Phys.\ Rev.\ D {\bf 93} (2016) no.7,  075009
  [arXiv:1506.05130 [hep-ph]].

\bibitem{Arhrib:2016rlj}
  A.~Arhrib, R.~Benbrik, S.~J.~D.~King, B.~Manaut, S.~Moretti and C.~S.~Un,
  arXiv:1607.08517 [hep-ph].



\bibitem{Aguilar-Saavedra:2013wba}
  J.~A.~Aguilar-Saavedra,
  EPJ Web Conf.\  {\bf 60} (2013) 16012
  [arXiv:1306.4432 [hep-ph]].

\bibitem{Barger:1995dd}
  V.~D.~Barger, M.~S.~Berger and R.~J.~N.~Phillips,
  Phys.\ Rev.\ D {\bf 52} (1995) 1663
  [hep-ph/9503204].

\bibitem{Frampton:1999xi}
  P.~H.~Frampton, P.~Q.~Hung and M.~Sher,
  Phys.\ Rept.\  {\bf 330} (2000) 263
  [hep-ph/9903387].

\bibitem{Barenboim:2001fd}
  G.~Barenboim, F.~J.~Botella and O.~Vives,
  Nucl.\ Phys.\ B {\bf 613} (2001) 285
  [hep-ph/0105306].

\bibitem{AguilarSaavedra:2002kr}
  J.~A.~Aguilar-Saavedra,
  Phys.\ Rev.\ {\bf D67} (2003) 035003
   Erratum: [Phys.\ Rev.\ D {\bf 69} (2004) 099901]
  [hep-ph/0210112].





\bibitem{Dawson:2012di}
  S.~Dawson and E.~Furlan,
  Phys.\ Rev.\  {\bf D86} (2012) 015021
  [arXiv:1205.4733 [hep-ph]].

\bibitem{Fajfer:2013wca}
  S.~Fajfer, A.~Greljo, J.~F.~Kamenik and I.~Mustac,
  JHEP {\bf 07} (2013) 155
  [arXiv:1304.4219 [hep-ph]].


\bibitem{Atre:2011ae}
  A.~Atre, G.~Azuelos, M.~Carena, T.~Han, E.~Ozcan, J.~Santiago and G.~Unel,
  JHEP {\bf 08} (2011) 080
  [arXiv:1102.1987 [hep-ph]].


\bibitem{ALEPH:2005ab}
  S.~Schael {\it et al.} [ALEPH and DELPHI and L3 and OPAL and SLD Collaborations and LEP Electroweak Working Group and SLD Electroweak Group and SLD Heavy Flavour Group],
  Phys.\ Rept.\  {\bf 427} (2006) 257
  [hep-ex/0509008].

\bibitem{Djouadi:2005gj}
  A.~Djouadi,
  Phys.\ Rept.\  {\bf 459} (2008) 1
  [hep-ph/0503173].

\bibitem{Fidalgo:2011ky}
  J.~Fidalgo, D.~E.~Lopez-Fogliani, C.~Mu\~noz and R.~Ruiz de Austri,
  JHEP {\bf 10} (2011) 020
  [arXiv:1107.4614 [hep-ph]].

\bibitem{Chen:2017hak}
  C.~Y.~Chen, S.~Dawson and E.~Furlan,
  arXiv:1703.06134 [hep-ph].













\bibitem{confnote}
  ATLAS collaboration, 
  ATLAS-CONF-2016-101.

\bibitem{Cacciari:2008gp}
  M.~Cacciari, G.~P.~Salam and G.~Soyez,
  JHEP {\bf 04} (2008) 063
  [arXiv:0802.1189 [hep-ph]].

\bibitem{Feldman:1997qc}
  G.~J.~Feldman and R.~D.~Cousins,
  Phys.\ Rev.\ {\bf D57} (1998) 3873
  [physics/9711021 [physics.data-an]].

\bibitem{Read:2002hq}
  A.~L.~Read,
  J.\ Phys.\  {\bf G28} (2002) 2693.

\bibitem{Czakon:2011xx}
  M.~Czakon and A.~Mitov,
  Comput.\ Phys.\ Commun.\  {\bf 185} (2014) 2930
  [arXiv:1112.5675 [hep-ph]].

\bibitem{delAguila:2000rc}
  F.~del Aguila, M.~Perez-Victoria and J.~Santiago,
  JHEP {\bf 09} (2000) 011
  [hep-ph/0007316].




\end{thebibliography}
\end{document}